\documentclass[fleqn,usenatbib]{mnras}

\usepackage{newtxtext,newtxmath}
\usepackage[T1]{fontenc}

\DeclareRobustCommand{\VAN}[3]{#2}
\let\VANthebibliography\thebibliography
\def\thebibliography{\DeclareRobustCommand{\VAN}[3]{##3}\VANthebibliography}
\newcommand{\lta}{\lower 2pt \hbox{$\, \buildrel {\scriptstyle <}\over {\scriptstyle \sim}\,$}}
\newcommand{\gta}{\lower 2pt \hbox{$\, \buildrel {\scriptstyle >}\over {\scriptstyle \sim}\,$}}

\usepackage{graphicx}	% Including figure files
\usepackage{amsmath}	% Advanced maths commands
\usepackage{subfigure}
\usepackage{caption}
\title[FRB Propagation]{Transparency of Fast Radio Burst Waves in Magnetar Magnetospheres}
\author[Y. Qu, P. Kumar and B. Zhang]{
Yuanhong Qu$^{1,2}$\thanks{E-mail: yuanhong.qu@unlv.edu}
,
Pawan Kumar$^{3}$\thanks{E-mail: pk@astro.as.utexas.edu}
and
Bing Zhang$^{1,2}$\thanks{E-mail: bing.zhang@unlv.edu}
\\
$^{1}$Nevada Center for Astrophysics, University of Nevada, Las Vegas, NV 89154\\
$^{2}$Department of Physics and Astronomy, University of Nevada Las Vegas, Las Vegas, NV 89154, USA\\
$^{3}$Department of Astronomy, University of Texas at Austin, Austin, TX 78712, USA
}

\date{}

\pubyear{2022}

\begin{document}
\label{firstpage}
\pagerange{\pageref{firstpage}--\pageref{lastpage}}
\maketitle

\begin{abstract}
At least some fast radio bursts (FRBs) are produced by magnetars. Even though mounting observational evidence points towards a magnetospheric origin of FRB emission, the question of the location for FRB generation continues to be debated. One argument suggested against the magnetospheric origin of bright FRBs is that the radio waves associated with an FRB may lose most of their energy before escaping the magnetosphere because the cross-section for $e^\pm$ to scatter large-amplitude EM waves in the presence of a strong magnetic field is much larger than the Thompson cross-section. We have investigated this suggestion and find that FRB radiation traveling through the open field line region of a magnetar's magnetosphere does not suffer much loss due to two previously ignored factors. First, the plasma in the outer magnetosphere ($r \gta 10^9 \ $cm), where the losses are potentially most severe, is likely to be flowing outward at a high Lorentz factor $\gamma_p \geq 10^3$. Second, the angle between the wave vector and the magnetic field vector, $\theta_B$, in the outer magnetosphere is likely of the order of 0.1 radian or smaller due in part to the intense FRB pulse that tilts open magnetic field lines so that they get aligned with the pulse propagation direction. Both these effects reduce the interaction between the FRB pulse and the plasma substantially. We find that a bright FRB with an isotropic luminosity $L_{\rm frb} \gta 10^{42} \ {\rm erg \ s^{-1}}$ can escape the magnetosphere unscathed for a large section of the $\gamma_p-\theta_B$ parameter space, and therefore conclude that the generation of FRBs in magnetar magnetosphere passes this test.
\end{abstract}


%%%%%%%%%%%%%%%%%%%%%%%%%%%%%%%%%%%%%%%%%%%%%%%%%%

%%%%%%%%%%%%%%%%% BODY OF PAPER %%%%%%%%%%%%%%%%%%

\section{Introduction}
Fast radio bursts are millisecond-duration and extremely-high-brightness-temperature radio signals \citep{Lorimer07,Thornton13,CHIME19}. The detection of FRB 200428 from the Galactic magnetar SGR 1935+2154 during an X-ray burst \citep{CHIME/FRB2020,Bochenek2020,Li21,Mereghetti20} suggested that at least some FRBs are produced by magnetars \citep{Thompson95,Thompson96}. However, the location of FRB generation from the magnetar is subject to debate \citep{Lu20,Margalit20}. In general, there are two types of widely discussed models \citep[e.g.][]{Zhang2020}: the pulsar-like models invoking emission inside or just outside a magnetar magnetosphere \citep[e.g.][]{Kumar2017,Kumar&Bosnjak2020,Yang&zhang2018,Yang&Zhang21,Wadiasingh20,Lu20,Lyubarsky2020,Zhang2022,Wang2022} and GRB-like models invoking relativistic magnetized shocks far away from the magnetosphere \citep[e.g.][]{Lyubarsky2014,Metzger2019,Beloborodov2020,Margalit20,Sironi2021}. Recent observations of repeating FRB sources revealed many pulsar-like observational properties of FRB emission, such as swings of polarization angle \citep{Luo2020}, high degree of circular polarizations \citep{Hilmarsson21,Xu21}, as well as the very high event rates in individual sources with consecutive burst separations as short as milliseconds \citep{LiD2021,Xu21}. These offer strong supports to the magnetospheric origin of FRBs.

On the other hand, the magnetospheric origin of FRBs was questioned by \cite{Beloborodov2021} from a theoretical argument. The basic argument is the following: Due to the extremely high luminosity of the FRB emission, an FRB generated within a magnetar magnetosphere would have  the amplitude parameter of the electromagnetic waves \citep{Luan2014}
\begin{equation}
a = \frac{e E_w}{m_e c \omega} \gg 1.
\end{equation}
Here $E_w$ and $\omega$ are the amplitude of the oscillating electric field and the frequency of the waves, and $e$, $m_e$ and $c$ are electron charge, mass and speed of light, respectively. At a large enough radius where the background magnetic field strength $B$ (which falls off as $r^{-3}$ for a dipolar configuration) becomes smaller than $E_w \simeq B_w$ (which falls as $r^{-1}$), the large-amplitude-wave effect becomes significant; this transition radius is at $r=R_{\rm E} \sim (10^9-10^{10} \ {\rm cm})$.  Electrons at rest in the magnetosphere are accelerated to relativistic speeds by the waves and the cross-section for the electrons to scatter the FRB waves becomes enormously large. \cite{Beloborodov2021} found that for an FRB with isotropic luminosity $\gta 10^{42} \ {\rm erg \ s^{-1}}$, the waves would be scattered away and the burst is effectively choked without reaching the observer. He further argued that the accelerated electrons can produce gamma-ray photons, which produce more electron-positron pairs that would further enhance the opacity. He concluded that FRBs, at least the high-luminosity ones, should not be produced inside magnetar magnetospheres. This raised the curious inconsistency between theoretical arguments and observational data.

There were two assumptions made in the \cite{Beloborodov2021} work. First, particles are assumed to be at rest in the magnetosphere before the arrival of the FRB wave; Second, the wave vector $\vec k$ is assumed to be perpendicular to the magnetic field vector $\vec B$, so that $\cos\theta_B = {\hat k}\cdot{\hat B}=0$. This is effectively assuming that FRB waves need to penetrate through the closed field line region in order to be observed. In a dynamical environment where an FRB is generated, on the other hand, the plasma in a magnetar magnetosphere is likely not static, but stream relativistically outwards. This is especially the case in the open field line region of a magnetar through which the FRB pulse generated in the magnetic polar cap region is expected to travel though \citep{Lu20}. Also $\theta_B$ becomes much smaller than $\pi/2$ in such a configuration. Therefore, the two assumptions made by \cite{Beloborodov2021} need to be dropped, and these drastically alter the conclusion he drew regarding the escape of an FRB pulse through the magnetar magnetosphere.

In this paper, we re-investigate the propagation of FRB waves in a magnetar magnetosphere under the more realistic conditions relevant to magnetospheric FRB emission models. We show that FRBs can escape from magnetar magnetospheres. The paper is organized as follows.  In section \ref{sec:cross-section}, we first perform a detailed numerical calculation of the cross section for an electron, initially at rest, to scatter FRB waves, and the dependence of this cross-section on $\theta_B$.  In section \ref{sec:streaming}, we provide arguments that suggest that the plasma in the open field region of the outer magnetosphere is likely rapidly moving outwards and present an order-of-magnitude estimate of the bulk Lorentz factor of the plasma. In section \ref{sec:thetaB} we provide several geometrical and physical arguments to show that $\theta_B$ is indeed small for relevant FRB models. The optical depth for FRBs is calculated in section \ref{sec:tau}, which is shown to be $< 1$ for a large part of the $\gamma_p - \theta_B$ parameter space even for bright FRBs with $L_{\rm frb} \sim 10^{42} \ {\rm erg \ s^{-1}}$. The main conclusions, along with some discussion, are in section \ref{sec:conclusions}.

\section{Large-amplitude wave scattering cross section in a strong background magnetic field}
\label{sec:cross-section}

In this section, we numerically study in detail the large-amplitude wave scattering cross section of an electron in a strong background magnetic field, paying special attention to the $\theta_B$-dependence of the cross section. 

Consider a charged particle that moves in response of an FRB's large-amplitude waves and a background strong magnetic field.
It is convenient to work with particle's canonical momentum \citep{Kumar&Lu2020}
\begin{equation}
\vec p=\gamma m_e\vec v+\frac{q}{c}\vec A.
\end{equation}
in the electromagnetic field for solving the momentum equation
\begin{equation}
\frac{d(\gamma m_e\vec v)}{dt}= q\left[ {\vec E}_w + \vec v\times({\vec B}_w + \vec B)\right]
  \label{force-eq}
\end{equation}
where $q$ is charge of the particle, $\vec B$ is the background, static, magnetic field, and ${\vec E}_w$ \& ${\vec B}_w$ are the electric and magnetic fields associated with the EM wave.

\subsection{General case of an arbitrary direction of the background magnetic field}
Let us consider the general case of a background magnetic field that is oriented at an arbitrary angle with respect to the FRB wave vector. Consider a spherical coordinate system with $z$-axis defined as the $\vec k$-direction and assume that the unit vector of the background magnetic field is $\hat{e}_{\rm bg}=(\sin\theta_B\cos\phi_B,\sin\theta_B\sin\phi_B,\cos\theta_B)$ in this coordinate system. There are infinite number of vector potentials to any pairs of ${\vec E}$ \& ${\vec B}$. We use this gauge freedom, and consider two different vector potentials for the static background $B$-field for calculating the conserved momenta
\begin{equation}
\vec A_1=B\sin\theta_B\sin\phi_Bz\hat{x}+B\cos\theta_Bx\hat{y}-B\sin\theta_B\cos\phi_Bz\hat{y},
\end{equation}
and
\begin{equation}
\vec A_2=B\sin\theta_B\sin\phi_Bz\hat{x}-B\cos\theta_By\hat{x}-B\sin\theta_B\cos\phi_Bz\hat{y},
\end{equation}
where $B$ is the amplitude of the static magnetic field. The vector potential for the FRB waves can be written as
\begin{equation}
\vec A_w=-\frac{cE_w}{\omega}\cos\psi\hat{x},
\end{equation}
where $\psi=kz-\omega t$ is wave phase, $k$ is wave number, $\omega$ is angular frequency, and ${\vec E}_w = -\partial \vec A_w/c\partial t$ is the electric field. For the first vector potential, $\vec A_1$, the Lagrangian is independent of the $y$ variable. Therefore the $y$-component of the canonical momentum is conserved
\begin{equation}
\gamma m_ev_y-\frac{q}{c}(B\sin\theta_B\cos\phi_Bz-B\cos\theta_Bx)={\rm const}.
  \label{vy1}
\end{equation}
We take the initial conditions as $x=y=z=0$ and $\vec v=0$ at $t=0$, i.e. the particle is at rest before it encounters the wave. The constant is zero for this initial condition, and the $x$-component of velocity can be written as
\begin{equation}\label{gv_y}
\gamma v_y=\omega_B(\sin\theta_B\cos\phi_Bz-\cos\theta_Bx).
\end{equation}
For the vector potential $\vec A_2$, the Lagrangian is independent of $x$. Therefore, the $x$-component of momentum is conserved
\begin{equation}
\gamma v_x-\frac{qE_w}{m_e\omega}\cos\psi+\frac{q}{m_ec}(B\sin\theta_B\sin\phi_Bz-B\cos\theta_By)={\rm const}.
 \label{vx1}
\end{equation}
The $x$-component of velocity can be written as
\begin{equation}\label{gv_x}
\gamma v_x=ac(\cos\psi-\cos\psi_0)-\omega_B(\sin\theta_B\sin\phi_Bz-\cos\theta_By),
\end{equation}
where $\psi_0$ is the initial phase of the FRB wave. The $z$-component of the equation of motion can be written as
\begin{equation}\label{zmotion}
\begin{aligned}
&\frac{d(\gamma m_ev_z)}{dt}=\frac{q}{c}[E_z+\vec v\times(\vec B_w+\vec B)]_z\\
&=\frac{q}{c}[v_xB_w\sin\psi+v_xB\sin\theta_B\sin\phi_B-v_yB\sin\theta_B\cos\phi_B],
\end{aligned}
\end{equation}
where $E_z=0$ because the wave-vector of the transverse FRB wave is along the $z$-axis. The kinetic energy equation of the particle can be written as
\begin{equation}\label{kinetic}
\frac{d(\gamma m_ec^2)}{dt}=qv_xE_w\sin\psi.
\end{equation}
Combining Equations (\ref{zmotion}) and (\ref{kinetic}), one can write the differential equation as
\begin{equation}
\frac{d}{dt}(\gamma v_z-\gamma c)=\omega_Bv_x\sin\theta_B\sin\phi_B-\omega_Bv_y\sin\theta_B\cos\phi_B,
  \label{vz1}
\end{equation}
where $\omega_B=eB/(m_ec)$ is the cyclotron frequency.
Since, $dx/dt=v_x$, $dy/dt=v_y$, the $z$-component of velocity can be written as
\begin{equation}\label{gv_z}
\gamma v_z=c(\gamma-1)+\omega_B\sin\theta_B(x\sin\phi_B-y\cos\phi_B),
\end{equation}
where we have used the initial conditions mentioned above.

Adding the squares of Equations (\ref{gv_x}), (\ref{gv_y}) and (\ref{gv_z}) and making use of the relation $\gamma^2v^2=c^2(\gamma^2-1)$, where $v=\sqrt{v_x^2+v_y^2+v_z^2}$, we obtain an explicit expression for the particle Lorentz factor
\begin{equation}
\begin{aligned}
&2\gamma[1-\omega_B\sin\theta_B(x\sin\phi_B - y\cos\phi_B)/c]=2+\frac{\omega_B^2}{c^2}\times\\
&(\sin\theta_B\cos\phi_Bz-\cos\theta_Bx)^2 + \frac{\omega_B^2 \sin^2\theta_B}{c^2}(x \sin\phi_B - y \cos\theta_B)^2 \\
& - 2\omega_B\sin\theta_B(x\sin\phi_B-y\cos\phi_B)/c\\
&+[a(\cos\psi-\cos\psi_0)-\omega_B(\sin\theta_B\sin\phi_Bz-\cos\theta_By)/c]^2.
\end{aligned}
\end{equation}

Particle trajectory is calculated using the first order differential equations (\ref{gv_y}), (\ref{gv_x}) and (\ref{gv_z}), and detailed numerical results will be presented in \S{\ref{numerical}}.

The cross section defined in the lab frame can be written as
\begin{equation}
\sigma=\frac{P}{S}=\frac{P}{E_w^2c/8\pi}=\frac{8\pi e^2P}{a^2m_e^2c^3\omega^2},
\label{eq:sigma}
\end{equation}
where the average radiation power $P$ emitted by a relativistic charged particle undergoing acceleration in the co-variant form is given by \citep{Jackson1998} 
\begin{equation}
\begin{aligned}
P&=\frac{2q^2}{3 c^3} \sum_{\alpha=0}^3 \frac{d(u^\alpha)}{ d\tau} \frac{d(u_\alpha)}{d\tau}\\
&=\frac{2q^2\gamma^2}{3c^3}\left[-\left(\frac{du^0}{dt}\right)^2+\sum_{i=x,y,z}\left(\frac{du^i}{dt}\right)^2\right],
\end{aligned}
\label{eq:P}
\end{equation}
where $u^\alpha=\gamma(c,v_x,v_y,v_z)$ is the four-velocity and $d\tau=dt/\gamma$ is the differential proper time, and its time derivative is given by equation (\ref{force-eq}). 

Since $P \propto \gamma^2$ (Eq.(\ref{eq:P})) and since $\gamma$ scales with $a$, the final $\sigma$ would be $\propto a^2$. Therefore, in our calculations sometimes we normalize $\sigma$ to $a^2 \sigma_{\rm T}$, where $\sigma_{\rm T}$ is the Thomson cross section.

\begin{figure*}
\begin{center}
\begin{tabular}{ll}
\resizebox{80mm}{!}{\includegraphics[]{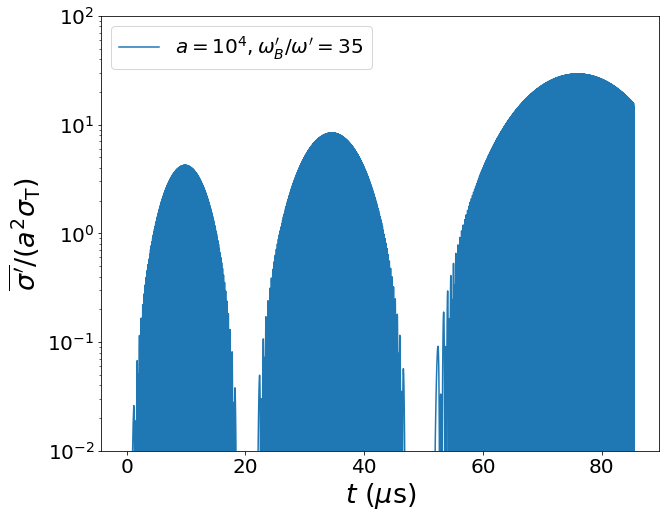}}&
\resizebox{80mm}{!}{\includegraphics[]{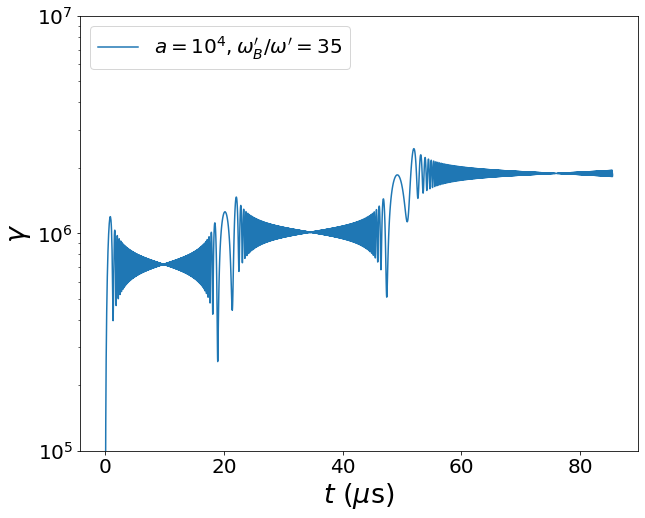}}
\end{tabular}
\caption{The normalized scattering cross section (left panel) and electron Lorentz factor (right panel) as a function of time.  Following parameters are adopted: $a=10^4$, $\theta_B'=10^{-1}$, $\phi_B=\pi/2$ and $\omega_B'/\omega'=35$.}
\label{Time}
\end{center}
\end{figure*}

\begin{figure}
	\includegraphics[width=\columnwidth]{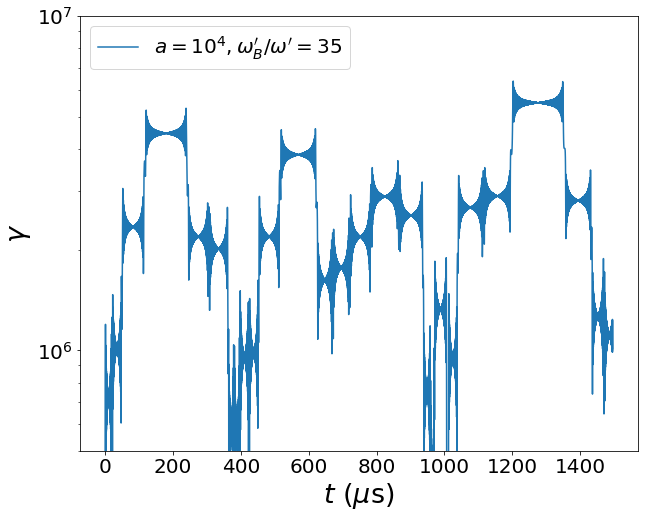}
    \caption{ The electron Lorentz factor as a function of time in a longer numerical run lasting for 1.4 ms.  Following parameters are adopted: $a=10^4$, $\theta_B'=10^{-1}$, $\phi_B=\pi/2$ and $\omega_B'/\omega'=35$.}
    \label{longer}
\end{figure}

\subsection{Numerical Results}\label{numerical}
In this section, we present our numerical results by solving the electron motion equation in a static background magnetic field and  linearly polarized large-amplitude EM waves. Starting from this subsection, we denote all the quantities as primes to specify that they are the values in the electron comoving frame.

\begin{figure}
	\includegraphics[width=\columnwidth]{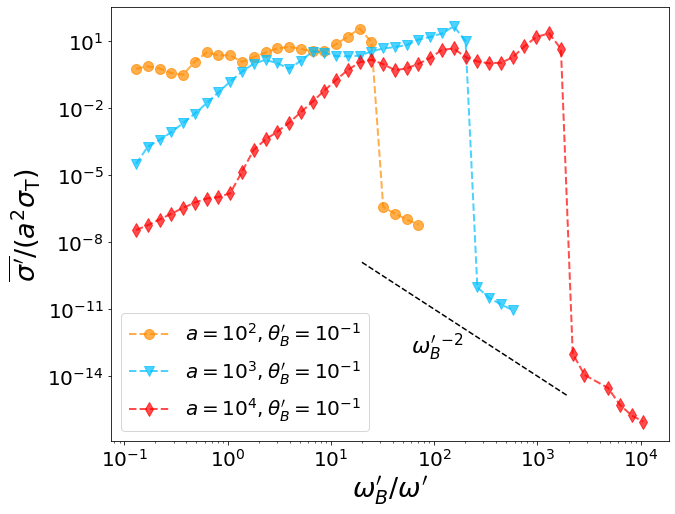}
	\includegraphics[width=\columnwidth]{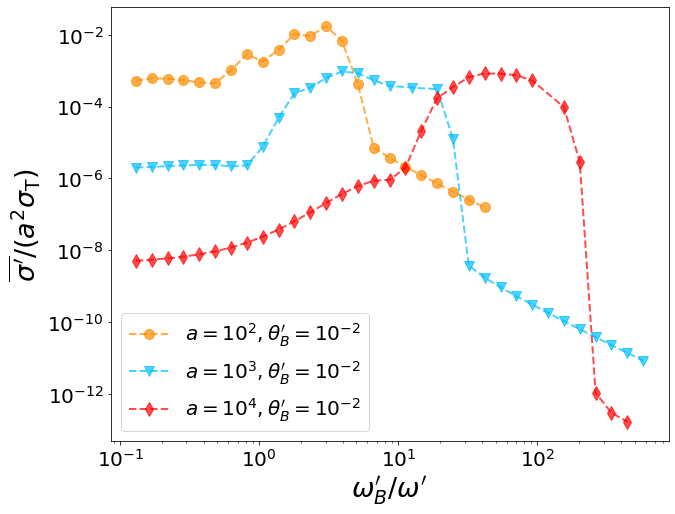}
	\includegraphics[width=\columnwidth]{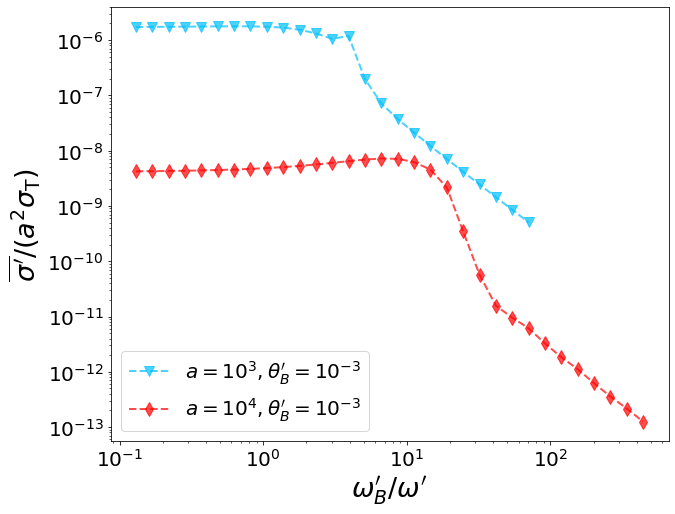}
	\caption{The normalized cross section $\overline{\sigma'}/(a^2\sigma_{\rm T})$ as a function of $\omega'_B / \omega'$ for different $a$ and $\theta'_B$ values. Upper, middle, and lower panels are for $\theta'_B=10^{-1}, 10^{-2}, 10^{-3}$, respectively. Orange, blue and red curves are for $a=10^2, 10^3, 10^4$, respectively. The black dotted line in the upper panel denote the $\propto (\omega'_B/\omega')^{-2}$-dependence.}
	\label{f123}
\end{figure}

\begin{figure}
	\includegraphics[width=\columnwidth]{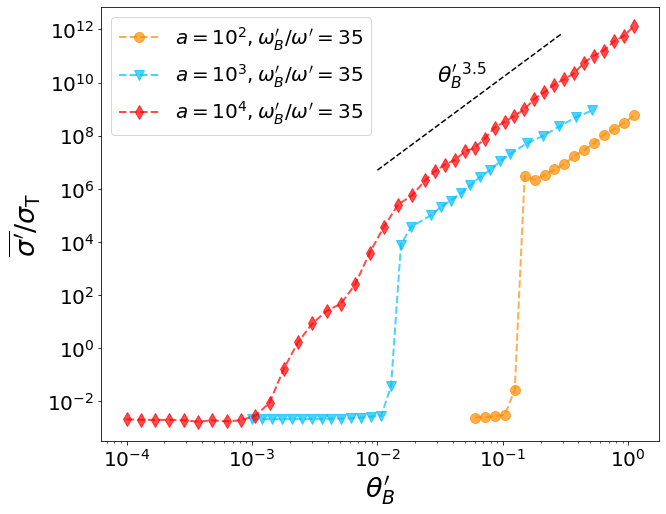}
    \caption{The normalized cross section $\overline{\sigma'}/\sigma_{\rm T}$ as a function of $\theta'_B$. The ratio $\omega_B'/\omega'$ is fixed to 35. Orange, blue and red curves are for $a=10^2, 10^3, 10^4$, respectively. The black dotted line represents the $\propto {\theta'_B}^{3.5}$-dependence.}
    \label{omega_B=35}
\end{figure}

We calculate the cross section as a function of the ratio of the cyclotron and FRB angular frequencies, i.e. $\omega_B'/\omega'$, and normalize it by $a^2 \sigma_{\rm T}$.
Note that $\sigma'$ and electron Lorentz factor vary significantly with time as shown in Fig. \ref{Time}. For a given set of parameters, one needs to run the simulations long enough to derive a reliable average value $\bar\sigma'$. In our numerical results, our  $\bar\sigma'$ values are typically derived for a time duration $t \sim 10^6 \omega_{\rm FRB}^{-1} \sim (2\pi)^{-1} \nu_{\rm GHz}^{-1}$ ms. { This timescale is much shorter than the FRB duration. The reason for choosing such a short timescale is that the result no longer changes significantly in a longer run so that one can use a shorter integration time to reduce  unnecessary expensive computations. In Figure \ref{longer}, we present a run for a much longer time duration of $t\simeq1.4$ ms, which is of order the typical FRB duration. One can see that the Lorentz factor does not continue to increase with time. Incidentally, the electron Lorentz factor attained in about 80 $\mu$s sometimes already exceeds the radiation reaction limit, which would further limit reachable Lorentz factor of the accelerated particles \citep{Beloborodov2021}. We note that radiation reaction is not considered in our calculations, so our calculated cross section can be regarded as a conservative upper limit. }

The numerical results of $\bar\sigma'/(a^2 \sigma_{\rm T})$ as a function of $\omega'_B/\omega'$ is presented in Figure \ref{f123}. The three panels show three values of $\theta'_B = 10^{-1}, 10^{-2}, 10^{-3}$, respectively. For $\theta_B'=10^{-1}$ (upper panel), one can see that the normalized cross section increases initially with $\omega_B/\omega$ and enters a plateau (with a maximum of $\sim{10}$) until a transition point, after which the cross section drops significantly to be $\ll 1$. At the end of each cross section curve, all the lines are aligned with each other and decrease in the same slope $\sim -2$ as $\omega'_B/\omega'$ increases. 
In the middle panel of Fig. \ref{f123} ($\theta_B'=10^{-2}$), the normalized cross section becomes systematically smaller than the upper panel. It also increases initially with $\omega_B'/\omega'$ and decreases after a transition point. In the lower panel of Fig. \ref{f123} ($\theta_B'=10^{-3}$), the normalized cross section is even smaller compared with the cases of larger angles. It stays nearly flat with $\omega'_B/\omega'$ before the rapid decline at the transition point. As seen in Figure \ref{f123}, the transition point to a much smaller cross section occurs at
\begin{equation}
    \omega'_B / \omega' \sim a \theta'_B.
    \label{eq:transition}
\end{equation}
We will further discuss the transition point and the corresponding critical radius in \S\ref{sec:tau}.

The numerical results of $\bar\sigma'/(a^2 \sigma_{\rm T})$ as a function of $\theta'_B$ is presented in Figure \ref{omega_B=35}, with $\omega_B'/\omega'=35$ fixed. 
The cross section is nearly flat at a small $\theta_B'$ but increases rapidly after the transition point defined by Eq.(\ref{eq:transition}), i.e. $\theta'_B \sim\omega_B'/(a\omega)$. At larger $\theta'_B$, we find that the cross section numerically follows an empirical relation
\begin{equation}
    \sigma' \propto {\theta'_B}^{3.5}.
    \label{eq:sigma-thetaB}
\end{equation}

\subsection{Scattering cross section for a moving electron}

The cross-sections calculated in previous sub-sections are for particles that were at rest before being hit by the FRB pulse. More generally and as discussed in following sections, the electrons are likely streaming out from the magnetosphere so that the electrons carry an initial Lorentz factor, $\gamma_0$, before being hit by the FRB pulse. One can derive the lab-frame cross section $\sigma$ for a moving electron using
the relationship with respect to the ``comoving frame'' cross-section $\sigma'$ using ({see Appendix \ref{A} for a derivation})
\begin{equation}
\sigma = \sigma' (1-\beta\cos\theta_B),
\label{eq:sigma}
\end{equation}
{where $\theta_B$ is the angle between incident wave and background magnetic field.} 

\begin{figure*}
\begin{center}
\begin{tabular}{ll}
\resizebox{80mm}{!}{\includegraphics[]{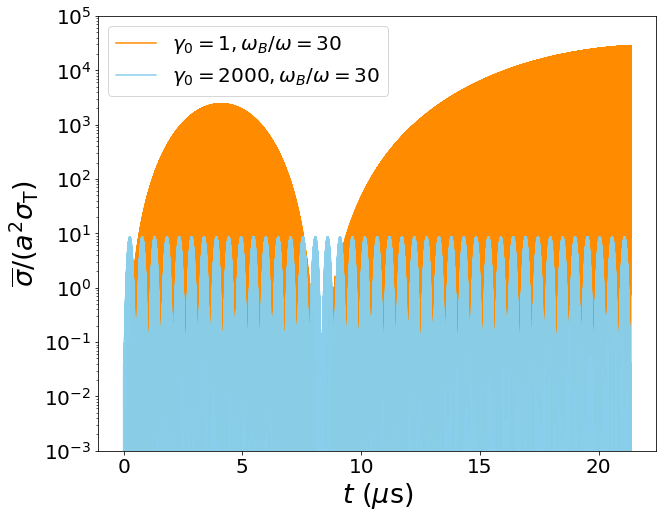}}&
\resizebox{80mm}{!}{\includegraphics[]{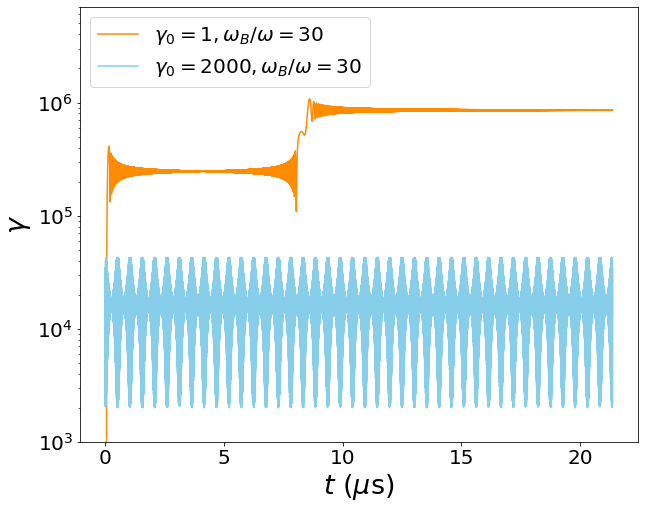}}
\end{tabular}
\caption{The normalized scattering cross section (left panel) and electron Lorentz factor (right panel) as a function of time for different initial LF ($\gamma_0=1$ and 2000, respectively) of electrons. Following parameters are adopted: $a=10^4$, $\theta_B=\pi/2$, $\phi_B=\pi/2$ and $\omega_B/\omega=30$.}
\label{fig:gamma_0}
\end{center}
\end{figure*}

One can also directly calculate the scattering cross-section for a moving particle without performing the above-mentioned transformation.
This can be done in a straightforward manner by changing the initial condition for particle Lorentz factor from 1 (as was done in \S\ref{sec:cross-section}) to $\gamma_0$ with the velocity-vector pointing along the background magnetic field. We can determine the particle four-velocity for this new initial condition using Equations (\ref{vy1}), (\ref{vx1}) and (\ref{vz1}), and calculate $\sigma$ as described before. The explicit expression for the particle Lorentz factor for the X-mode FRB radiation, when the magnetic field is perpendicular to the FRB pulse propagation direction, is
\begin{equation}
\begin{aligned}
&\gamma = \gamma_0 + \frac{ [a(\cos\psi-\cos\psi_0)-\omega_B z/c]^2 + \omega_B^2 x^2/c^2}{2(\gamma_0 - \omega_B x/c)},
\end{aligned}
\end{equation}\label{eq:gam-gam0}
This equation shows that the particle Lorentz factor ($\gamma$) is smaller when the initial Lorentz factor of the particle ($\gamma_0$) is much greater than 1 as shown in Fig. \ref{fig:gamma_0} right panel. The increase of $\gamma$, i.e. $\gamma-\gamma_0$ in Eq.(\ref{eq:gam-gam0}), for a smaller $\gamma_0$ is much greater than that for a larger $\gamma_0$, when $a\gg\gamma_0$, since $x$ is of the order of the Larmor radius and $\omega_B x/c \sim \gamma$. The smaller value of $\gamma$ results in a smaller cross-section in the lab frame as shown in Fig. \ref{fig:gamma_0} left panel. The combined effect of $\theta_B < 1$ and $\gamma_0 > 1$ on the scattering cross-section is the same as the case of background magnetic field perpendicular to the EM pulse momentum (i.e. $\theta_B=\pi/2$) and the particle initial Lorentz factor along the background field of $\sim \gamma_0/\theta_B$, which leads to a large reduction in the interaction between FRB pulse and $e^\pm$. We should note that the component of particle 4-velocity along the direction of the background magnetic field is conserved when the radio pulse polarization is the X-mode. Thus, the memory of the initial condition of the plasma motion along the magnetic field is preserved even as the other components of the particle momentum oscillate and fluctuate wildly.

\section{Streaming relativistic plasma in the FRB propagation regions}\label{sec:streaming}
In this section, we show that three physical scenarios can accelerate the plasma in the open field line region of a magnetar magnetosphere to high Lorentz factors.

\subsection{Standard pulsar mechanism}\label{sec:pulsar}

Rotating magnetized neutron stars induce a large electric potential due to the unipolar effect, which can accelerate particles in the open field line regions. Even without an FRB, the standard pulsar mechanism is likely operating. Observations of SGR 1935+2154 showed that even before the FRB generating X-ray burst, there have been many X-ray bursts emitted that were not associated with FRBs \citep{Lin20}. These activities may reconfigure the magnetic structure of the star to facilitate the FRB generation later on. In the following, we give an estimate of the outflow Lorentz factor of the $e^+ e^-$ plasma in the open field line region based on the standard pulsar mechanism.

In the polar cap region near the magnetar surface, there is likely a region where the net change density $\rho < \rho_{\rm GJ}$, where $\rho_{\rm GJ} \sim -({\vec \Omega \cdot \vec B})/2\pi c$ is the Goldreich Julian density \citep{GJ1969} to maintain corotation of the magnetosphere. Such a charge-deficit region is called an inner gap, which may be produced if the surface ion binding energy is strong enough \citep{ruderman75} or, in a space-charge-limited flow due to flaring of the open field lines \citep{Arons79} or the frame-dragging effect in a general relativistic treatment \citep{Muslimov92}. The maximum electric potential due to the unipolar effect
\begin{equation}
\Phi_{\rm max}\sim \frac{B_\star R_\star^3\Omega^2}{2c^2}\simeq(6.6 \times10^{15} \ {\rm Volts}) \ B_{\star,15}R_{\star,6}^3P^{-2}
\end{equation}
usually cannot be achieved because the gap is screened by electron-positron pairs produced by $\gamma$-rays through $\gamma-B$ \citep{daugherty96,thompson08} or $\gamma-\gamma$ processes \citep{zhang2001} near the magnetar surface\footnote{Photon splitting is also discussed in the strong magnetic fields \citep{baring98}, but it would not suppress pair production completely because only one mode is allowed to split \citep{usov02}.}. The actual gap potential depends on the $\gamma$-ray radiation mechanism (curvature radiation or inverse Compton scattering), the type of the gap (vacuum or space-charge-limited flow), as well as the curvature radius of the near-surface magnetic field  \citep{Zhang2000}. The energies of the secondary pairs are, on the other hand, similar to each other to order of the magnitude. In the following, we adopt the simplest vacuum gap model and discuss curvature radiation as the $\gamma$-ray emission mechanism to perform the estimate. 

For a vacuum gap \citep{ruderman75}, the gap height $h$ is in principle the sum of three terms \citep{Zhang2000}, i.e. $h = l_{\rm acc} + l_{\rm CR} + l_\gamma$, where $l_{\rm acc}$ is the distance of  acceleration for the primary lepton to reach the characteristic Lorentz factor $\gamma_c$, 
\begin{equation}\label{l_CR}
l_{\rm CR}\simeq c\left(\frac{P_{\rm CR}}{\hbar\omega_{c}}\right)^{-1}=\frac{9\hbar\rho c}{4e^2\gamma_c}\simeq(3.1\times10^2 \ {\rm cm}) \ \rho_6\gamma_{c,6}^{-1}
\end{equation}
is the characteristic distance for the lepton to emit a $\gamma$-ray photon via curvature radiation ($P_{\rm CR}=2\gamma^4e^2c/3\rho^2$ is the curvature radiation power, and $\omega_c=3\gamma_c^3 c/2\rho$ is curvature radiation frequency), and 
\begin{equation}\label{l_pair}
\begin{aligned}
l_{\gamma,\rm CR}&=\frac{4.4}{(e^2/\hbar c)}\frac{\hbar}{m_ec}\frac{B_q}{B_\perp}{\rm exp}\left(\frac{4}{3\chi}\right)\\
&\simeq(1.03\times10^6 \ {\rm cm}) \ B_\perp^{-1}{\rm exp}\left(\frac{4}{3\chi}\right)
\end{aligned}
\end{equation}
is the mean free path of the $\gamma$-ray before producing $e^\pm$ in the strong magnetic field. Here $B_q=m_e^2c^3/(e\hbar)\simeq 4.4\times10^{13} \ \rm G$ is the critical magnetic field, 
\begin{equation}
  B_\perp=B\sin\theta_B  =\frac{h}{\rho}B
  \label{B_perp}
\end{equation}
is the perpendicular magnetic field component with respect to the photon propagation direction ($\theta_B \sim h/\rho$ is the angle between the photon and the $B$ field). The parameter $\chi$ is 
\begin{equation}\label{chi}
\chi_{\rm CR}=\frac{E_\gamma}{2m_ec^2}\frac{B_\perp}{B_c}=\frac{3\gamma_c^3c}{4\rho m_ec^2}\frac{B_\perp}{B_q},
\end{equation}
where $E_\gamma=\hbar\omega_c$ is the energy of the $\gamma$-ray photon. 
It turns out that in the vacuum curvature radiation model, $l_{\rm acc}$ is much smaller than $l_\gamma$ and $l_{\rm CR}$, so that $h \sim l_\gamma + l_{\rm CR}$, and the total potential across the gap is \citep{ruderman75}
\begin{equation}\label{Phi}
\Phi_{\rm tot,CR} \simeq \frac{\Omega B}{c}h^2\simeq\frac{\Omega B}{c} (l_\gamma+l_{\rm CR})^2.
\end{equation}
Numerically solving Equations (\ref{l_CR})-(\ref{Phi}) self-consistently, one gets the Lorentz factor of the primary leptons from the gap
\begin{equation}
\gamma_{\rm pri, CR}=\frac{e\Phi_{\rm tot}}{m_ec^2}\simeq3.4\times10^6.
\end{equation}
This gives the characteristic $\gamma$-ray energy
\begin{equation}
E_{\gamma,\rm CR}=\frac{3\gamma_{\rm pri}^3c\hbar}{2\rho}\simeq 
 (1.2 \ {\rm GeV})\ \rho_6.
\end{equation}
This energy is evenly split to the electron-positron pair so that each lepton gets $E_{\pm,\rm CR}=E_{\gamma,\rm CR}/2\simeq(0.6 \ {\rm GeV}) \ \rho_6$. The Lorentz factor of the secondary charged pair plasma is therefore
\begin{equation}
\gamma_{p,\rm CR}=\frac{E_{\rm pair}}{m_ec^2}\simeq1.2\times10^3 \rho_6.
\label{eq:gammap}
\end{equation}
These pairs flow outwards along the open field line regions with a relativistic speed. 

In the case of space-charge-limited models, the acceleration length $l_{\rm acc}$ would become larger. However, since both $l_{\rm CR}$ and $l_\gamma$ are not very long near the surface of the magnetar, it is expected that the estimated $\gamma_{p,\rm CR}$ would not be too different from that estimated in Eq.(\ref{eq:gammap}).

{ The primary particles may also radiate $\gamma$-ray photons via inverse Compton scattering (ICS). However, for a typical magnetar environment, the ICS process is not important near the surface. This is because resonant Compton scattering, which requires  the seed photon energy to be $\sim \omega_B/\gamma_e$, is in the Klein-Nishina regime, because $\gamma \hbar \omega_B > \gamma m_e c^2$ when $B > B_q = 4.414\times 10^{13}$ G. One can then write the typical ICS photon energy as \citep[e.g.][]{zhang2002}
\begin{equation}
\begin{aligned}
E_{\rm \gamma,ICS}& \simeq {\rm min}(\gamma_{\rm pri}^2kT,\gamma_{\rm pri}m_ec^2)\\
&={\rm min}(0.86\gamma_{\rm pri,3}^2T_7,0.51\gamma_{\rm pri,3}){\rm GeV}.
\end{aligned}
\end{equation}
Let us estimate the ICS power as $P_{\rm ICS}\sim (4/3)\gamma^2\sigma cU_{\rm ph}$ where $\sigma$ is suppressed from the KN-modified cross section $\sigma_{\rm KN} \lesssim \sigma_{\rm T}$ by another factor of $\sim (\omega/\omega_B)^2$ with $\omega \sim kT/(\hbar\gamma)$ and $U_{\rm ph} \sim aT^4$ is the surface thermal photon energy density. With typical values at the magnetar surface $kT \sim 5$ keV and $B \sim 10^{15}$ G, one finds that $P_{\rm ICS} \lesssim 4.2 \times 10^{-4} \ {\rm erg \ s^{-1}}$, which is $\ll P_{\rm CR} \sim 6.2\times 10^5 \ {\rm erg \ s^{-1}}$, so that the ICS process is much less efficient than the CR process to produce pairs. We can therefore ignore the ICS process and apply Eq.(\ref{eq:gammap}) to estimate $\gamma_p$. 
}

\subsection{Alfv\'en waves acceleration}\label{sec:Alfven}

The detection of many X-ray bursts before FRB 200428 from SGR J1935+2154 \citep{Lin20} suggests that it is likely that Alfv\'en waves with different amplitudes have been propagating in the magnetar magnetosphere before the FRB is launched. It is also possible that FRBs have precursors like we see for many GRBs. These Alfv\'en waves, which preceded the FRB pulse, would accelerate the plasma in the open field line regions and eventually escape the magnetosphere. In fact, a small segment at the head of the magnetic disturbance that produced the FRB pulse could accelerate and evacuate the plasma in the outer magnetosphere before the arrival of the radio pulse there. In any one of these scenarios, the FRB pulse is likely to propagate through highly evacuated outer magnetosphere, and thereby suffer little scattering or loss of energy.

It is only the highly luminous FRB radio waves, with $L\gta 10^{42}$ erg s$^{-1}$, that are scattered by $e^\pm$ with cross-section much larger than $\sigma_{\rm T}$ in the outer magnetosphere of a magnetar ($r\gta 10^9$cm), which might have difficulty escaping intact. The magnetic disturbance that produces these high luminosity waves have amplitude at the surface of the NS, $\delta B\sim 10^{11}$G assuming an efficiency of a few percent for converting magnetic energy to radio waves. The dimensionless wave amplitude at the surface is $\delta B/B \sim 10^{-4}$ for a magnetar field strength of 10$^{15}$G. Since magnetic disturbances follow the field lines, their amplitude decreases with radius as $r^{-3/2}$, and the dimensionless wave amplitude increases as $r^{3/2}$. Thus, the magnetic disturbance becomes nonlinear at $r\gta 10^9$, and that leads to the ejection of the plasma as suggested by MHD simulations of NS magnetosphere, e.g. \cite{chen2022}.

It should also be pointed out that the minimum particle density required to prevent charge starvation for an Alfven wave of amplitude $\delta B$ and transverse wave vector $k_\perp$ is $k_\perp \delta B/(8\pi q)$. At $r\sim 10^9$ cm, this density is $5 \times 10^8$ cm$^{-3}$ for Alfven waves of frequency 10 kHz, and $k_\perp$ of the order of the wave-vector at that height. The Goldreich-Julian density is $10^5$ cm$^{-3}$ at that radius for a magnetar with spin period of 1 s and surface magnetic field of 10$^{15}$G. Thus, the particle density needed to prevent charge starvation is $\sim 10^4 n_{\rm GJ}$. If the density were to be smaller, then the Alfven wave will develop a strong electric field parallel to the magnetic field that would accelerate particles to high Lorentz factors and eject the plasma along the open field lines before the arrival of the FRB pulse. 

\subsection{Ponderomotive force acceleration}\label{sec:ponder}
A third acceleration mechanism is associated with the ponderomotive force of the FRB waves themselves. A ponderomotive force is the force due to the gradient of energy in an electromagnetic wave packet. An FRB has a short duration and naturally develops a huge ponderomotive force at the front end of the wave pulse.
With a large amplitude of $a \gg 1$, the ponderomotive force is in the relativistic regime and can be written as \citep{Bauer1995,Yang&Zhang2020}
\begin{equation}
\vec F_p=-m_e c^2\nabla\left(1+\left< {a}^2 \right> \right)^{1/2} \simeq -m_e c^2 (\nabla a),
\end{equation}
where $\vec a={e\vec A_w}/({m_ec^2})=-{eE_w}/({m_ec\omega})\cos\psi\hat{x}$, and $\Vec{A}_w$ is the potential vector of FRB waves defined in \S\ref{sec:cross-section}, and $a = | \vec a| \gg 1$ has been adopted in the second half of the equation. An order of magnitude estimate suggests that an electron in front of the FRB pulse can reach a Lorentz factor $\gamma \sim (m_e c^2) (a / \Delta r) (\Delta r) / (m_e c^2) \sim a$. Since $a \gg 1$ for a typical FRB, the plasma in front of the FRB packet is accelerated to $\gamma \gg 1$, and carried with the FRB  pulse outside the magnetosphere even if the two mechanisms discussed in Sections \ref{sec:pulsar} and \ref{sec:Alfven} do not operate. 

\section{Angle between FRB waves and the magnetic field lines}
\label{sec:thetaB}

In this section, we discuss $\theta_B$ in realistic FRB models and show that it is typically small.

\subsection{Dipolar magnetic field geometry}
\label{dipole-geometry}

The near-surface magnetosphere of a magnetar likely has a complex multi-polar configuration. However, at large radii in the magnetosphere ($r\gta 10^9$ cm$^{-3}$), where scattering cross-sections might become very large, the field configuration is likely dominated by the dipolar component. Considering that an FRB radio emission is generated in the open field line region of the magnetar \citep[e.g.][]{Lu20}, { we perform an order of magnitude estimate of $\theta_B$ assuming a dipolar geometry.}

\begin{figure*}
\begin{center}
\begin{tabular}{ll}
\resizebox{80mm}{!}{\includegraphics[]{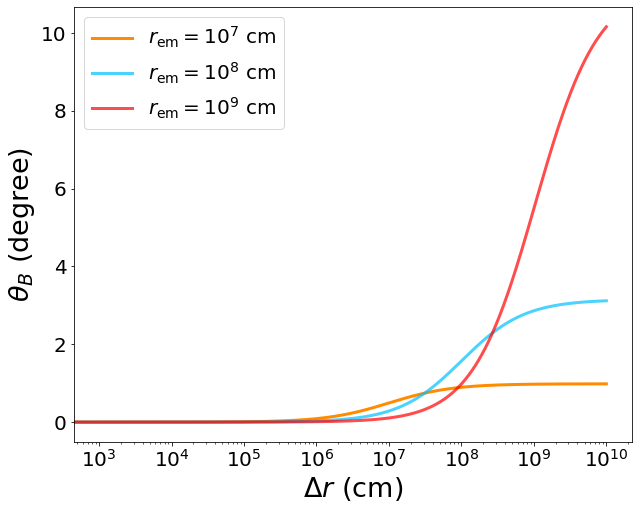}}&
\resizebox{80mm}{!}{\includegraphics[]{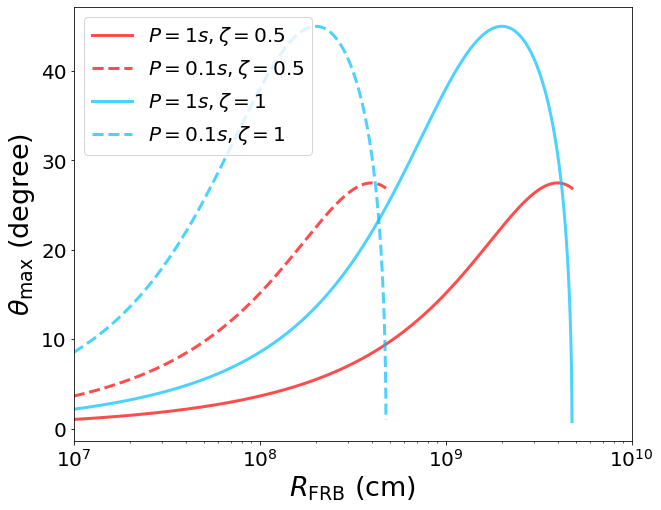}}
\end{tabular}
\caption{The $\theta_B$ values for different parameters in the dipolar geometry. Left panel: $\theta_B$ as a function of the wave propagation distance $\Delta r$. Three emission radii $r_{\rm em} = 10^7 \ \rm cm$ (orange solid line), $10^8 \ \rm cm$ (blue solid line) and $10^9 \ \rm cm$ (red solid line) are considered. Other parameters: magnetar period $P=1 \ \rm s$, radius $R_\star=10^6 \ \rm cm$, and $\zeta=0.5$. 
Right panel: the maximum $\theta_{\rm B,max}$ value as a function of  $r_{\rm em}$.  The red and blue lines stand for $\zeta = 0.5$ and $\zeta=1$ and the solid and dashed lines stand for $P=1$ s and $P=0.1$ s, respectively.}
\label{theta_B}
\end{center}
\end{figure*}

Let us consider a star-centered magnetic dipole
and that the FRB emission point is at $(x_0,y_0)$, with
\begin{equation}
x_0=\frac{R_\star \sin^2\theta_0}{\sin^2(\zeta \theta_p)} \  \  \ {\&} \ \ \ y_0=\frac{R_\star \sin^2\theta_0\cos\theta_0}{\sin^2(\zeta \theta_p)},
\end{equation}
where $\theta_0$ is the polar angle of the emission point, $R_\star$ is the magnetar radius, $\theta_p=\arcsin\sqrt{R_\star/R_{\rm LC}}$ is the polar cap opening angle and the parameter $0<\zeta<1$ characterizes a field line in the open field line region.
The FRB wave vector is tangential to the field line at the emission point and increases moderately as it propagates outwards. 
Let us define $\theta_\mu$ as the angle between the direction of the magnetic field and the magnetic axis. Its dependence on the polar angle $\theta$ is \citep{Qiao&Lin1998}:
$\tan\theta_{\mu}={3\tan\theta}/({2-\tan^2\theta})$. For FRB waves emitted from $(x_0, y_0)$ (with $\theta_{\mu,0}$) and traveling to an arbitrary point $(x, y)$ (with $\theta_{\mu}$), the angle between the wave vector and the local $B$ vector is
\begin{equation}
\theta_B=\theta_{\mu}-\theta_{\mu,0}=\arctan\frac{3\tan\theta}{2-\tan^2\theta}-\arctan\frac{3\tan\theta_0}{2-\tan^2\theta_0}.
\end{equation}
We present $\theta_B$ as a function of FRB wave propagation distance $\Delta r$ for different emission radii in Fig. \ref{theta_B} left panel. The angle $\theta_B$ increases with emission radius. For $r_{\rm em}=10^8 \ \rm cm$ and $\zeta=0.5$, the maximum $\theta_B$ is a few degrees or $\sim 0.1$ radian.

The angle $\theta_B$ reaches the maximum value in the magnetosphere at the light cylinder radius. For a dipolar geometry, 
This maximum angle $\theta_{B,\rm max}$ can be written as
\begin{equation}
\theta_{B,\rm max}=\cos^{-1} \left\vert \frac{1+k_{\rm em}k_{\rm lc}}{\sqrt{1+k_{\rm em}^2}\sqrt{1+k_{\rm lc}^2}} \right\vert.
\end{equation}
In the right panel of Fig. \ref{theta_B} we plot the maximum $\theta_{B}$ encountered by the FRB waves as a function of the emission radius $r_{\rm em}$ for different field lines and different rotation periods $P$. One can see that a larger emission radius and a shorter rotation period $P$ tend to make a larger $\theta_{B,\rm max}$. In any case, for emission radius for FRB radio waves of $r_{\rm em} \lta 10^8$ cm, $\theta_{B,\rm max}$ is { typically smaller} than $10^{\rm o}$ for a large parameter space. 

{ In reality, the magnetic field configuration would deviate from the dipolar geometry near the light cylinder, where the toroidal $B$ field increases significantly so that $\theta_B$ may approach $90^{\rm o}$. Also the mis-alignment between the spin and magnetic axes would introduce further complications near the light cylinder. On the other hand, since the background magnetic field $B_{\rm bg}$ drops more rapidly than $B_w$ with $r$, near the light cylinder $B_{\rm bg}$ is dynamically  insignificant. The $\vec k-\vec B$ alignment effect as discussed in Section \ref{sec:alignment} below becomes more significant, so that the geometric estimation of $\theta_B$ is no longer relevant. In any case, the calculation presented here gives a reasonable order-of-magnitude estimate for $\theta_B$ in the inner magnetosphere where $B_{\rm bg}$ is still not much smaller than $B_w$. } 

\subsection{Alignment between $\vec B$ and $\vec k$ vectors}\label{sec:alignment}

The calculation of $\theta_B$ in the previous sub-section assumed a dipole magnetic field geometry. In reality, it is likely that the strong radio pulse of an FRB moving outward would modify the magnetic field configuration and induce an alignment between the $\vec B$ and $\vec k$ vectors. Even though approving this requires numerical simulations, in the following we present an analytical argument to justify it.

As shown in \S\ref{dipole-geometry}, there is a high probability that the EM pulse will find itself in the region of the magnetosphere at $r>10^9$ cm where magnetic field lines are open, i.e. one end of the magnetic field line is tied to the NS surface and the other end extends beyond the light cylinder. In this case, the strong EM pulse is likely to force the plasma to move radially outward along the wave-vector of the wave, and tilt the background magnetic field orientation so that it is aligned with the wave-vector ${\vec k}$. The indication for this behavior is via the following consideration and construction. Suppose that instead of this scenario what the EM pulse does is to sweep up the plasma and force it to move with Lorentz factor $\gamma$, or speed $v$ along the ${\vec k}$-vector, and the background magnetic field lines are compressed and carried with the plasma pointing perpendicular to the ${\vec k}$-vector. Since the field lines are tied to the NS surface, the magnetic field cannot be perpendicular to the wave-vector everywhere. They must become highly curved in some region. In any case, in the region where the field lines are perpendicular to ${\vec k}$ and moving with the plasma at speed $v$, the electric field in the co-moving frame is zero. Therefore, the fields in the NS rest frame are obtained by Lorentz transformation
\begin{equation}
    {\vec B_\perp} = \gamma {\vec B_\perp'}, \quad {\vec E_\perp} = -\frac{\gamma}{c} \left( {\vec v \times B_\perp'}\right), \quad {\vec E}_\parallel = 0.
\end{equation}
where the parallel components of the field refers to the direction along ${\vec k}$ of the EM pulse. The momentum density in the swept up magnetic field in the NS rest frame is
\begin{equation}
{\vec P}_B = \frac{ {\vec E\times \vec B}}{4\pi c} = \frac{1}{4\pi c} \left[ \gamma^2 B_\perp'^2\, \frac{{\vec v}}{c} - \frac{\gamma v B_\parallel}{ c} {\vec B'_\perp} \right].
\end{equation}
The first term in the momentum density is along ${\vec v}$ (same as the direction of ${\vec k}$-vector), and that is fine as some fraction of the momentum of the EM pulse is transferred to the plasma and the magnetic field. However, the second term in the momentum density is perpendicular to ${\vec k}$, if $B_\perp\not=0$ and $B_\parallel\not=0$, which cannot be avoided, at least during the transition time when the magnetic field orientation is being forced to become perpendicular to ${\vec k}$. The magnitude of this perpendicular momentum component is of the order of the component along ${\vec k}$ when $\gamma\sim 1$. Moreover, the total magnitude of these momentum components are of the order of the momentum carried by the EM pulse in the region where the NS magnetic field is not much smaller than the electric field associated with the EM pulse, and when the EM pulse is scattered efficiently in the medium. The momentum is transferred from the EM pulse to the plasma which then transfers it to the swept-up field. The problem is that according to numerical calculations of particle dynamics under the combined forces of the EM pulse and the background magnetic field (\S3), the momentum kick imparted to particles perpendicular to ${\vec k}$ oscillates with time and the time averaged value of this component is small compared with the momentum imparted along ${\vec k}$ by at least a factor of 10 when the angle between the background magnetic field and ${\vec k}$ is of the order of 1 radian, i.e. there is insufficient momentum transfer taking place from the EM pulse to plasma to account for the total momentum in the swept-up fields perpendicular to ${\vec k}$. The way out of this contradiction is that the EM pulse does not cause the background magnetic field to become perpendicular to ${\vec k}$ but rather forces the field to become nearly parallel to ${\vec k}$, which requires much smaller momentum transfer.

\subsection{Transformation to the plasma co-moving frame}
\begin{figure}
	\includegraphics[width=\columnwidth]{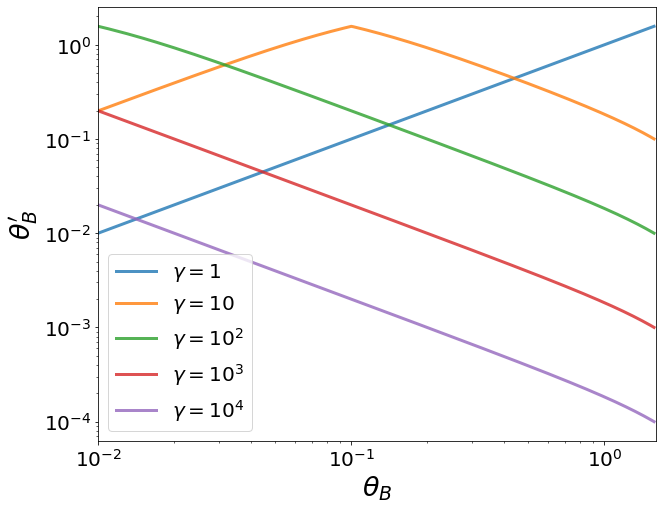}
    \caption{The relation between the $\theta'_B$ and $\theta_B$ for different plasma Lorentz factor $\gamma$ values.}
    \label{theta_B_prime}
\end{figure}

In order to make use of the cross section calculation results studied in \S\ref{sec:cross-section}, one needs to get to the rest frame of an electron before it is hit by the FRB radio pulse (the primed frame). Let us consider that the plasma encountered by the EM wave is moving along the magnetic field with Lorentz factor $\gamma_p$. The plasma co-moving frame angle between the magnetic field and the radio pulse propagation direction, $\theta'_B$, is related to the angle $\theta_B$ in the lab frame through
\begin{equation}
\sin\theta'_B = (\omega/\omega') \sin\theta_B,
\end{equation}
where 
\begin{equation}
\frac{\omega}{\omega^{'}} = {\cal D} = \frac{1}{\gamma_p(1 - \beta_p\cos\theta_B)} \approx \frac{2\gamma_p}{\left[ 1 + (\gamma_p\theta_B)^2 \right]}
\label{w-prime}
\end{equation}
is the Doppler factor (defined in Eq.(\ref{eq:Doppler})) that connects  quantities from the co-moving frame to the observer frame.
The last approximate expression is valid for $\theta_B \ll 1$ and $\gamma_p \gg 1$. We can rewrite the angle in the co-moving frame as follows 
\begin{equation}
\sin\theta'_B ={\cal D} \sin\theta_B \quad {\rm or} \quad \theta'_B \approx \frac{2}{\gamma_p \theta_B}
\label{thetaB-prime}
\end{equation}
for $\gamma_p \theta_B\gg 1$. One can see that $\theta'_B$ is smaller than $\theta_B$ when
$\gamma_p > 2\theta_B^{-2}$. In general, the relationship between $\theta'_B$ and $\theta_B$ are presented in Fig. \ref{theta_B_prime}.
One can see that $\theta_B'$ could be very small even if $\theta_B$ is large if $\gamma_p \gg 1$.

\section{Scattering optical depth for FRBs generated in a magnetar magnetosphere}\label{sec:tau}
With the preparation from the previous sections, we are now in the position to calculate the scattering optical depth for an FRB generated in a magnetar magnetosphere. We normalize FRB radio luminosity to $L_{\rm frb}=(10^{42} \ {\rm erg \ s^{-1}}) \ L_{\rm frb,42}$, which is typical for a bright non-repeating FRB and more than an order of magnitude larger than what we observe for repeating FRBs. The EM wave nonlinear parameter is given by
\begin{equation}
a=\frac{e E_w}{m_ec\omega}=\frac{e L_{\rm frb}^{1/2}}{m_e c^{3/2}\omega  r}\simeq1.6\times10^4 \ L_{\rm frb,42}^{1/2}\nu_9^{-1}r_9^{-1},
\end{equation}
where we consider the typical FRB frequency $\nu=(10^9 \ {\rm Hz}) \ \nu_9$ and normalize the radius to $r=(10^9 \ {\rm cm}) \ r_9$. 

Before writing down the general expression of the scattering optical depth of the FRB, it is informative to introduce 5 characteristic radii:
\begin{itemize}
\item The FRB emission radius $R_{\rm FRB}$ (which is the same as $r_{\rm em}$ defined in Section \ref{sec:thetaB}): This is the radius where the FRB radio waves are generated. According to the published magnetospheric FRB models, this radius is likely 10s-100s of NS radii \citep[e.g.][]{Kumar&Bosnjak2020,Lu20,Zhang2022}, which is $\sim (10^7 - 10^8)$ cm.
\item The inner transition radius $R_{\rm E}$:
As long as the electric field of the EM pulse $E_w$ is weaker than the background magnetic field $B_{\rm bg}$, the cross-section for the scattering of the pulse by $e^\pm$ is small, which is of the order $\sim \sigma_{\rm T} (\omega'/\omega'_B)^2$ for the EM X-mode photons. 
However, the cross section increases substantially when $E_w>B_{\rm bg}$, where $B_{\rm bg}$ is the background static B-field.
For a dipolar magnetic field with $B_\star=10^{15}$ G at the surface of the NS, this transition radius where $E_w=B_{\rm bg}$ is given by
\begin{equation}
R_{\rm E}=\left( \frac{B_\star R_{\star}^3 c^{1/2}}{L_{\rm frb}^{1/2}} \right)^{1/2} = (7.7\times10^8\,{\rm cm})\; B_{\star,15}^{1/2} R_{\star,6}^{3/2} L_{\rm frb,42}^{-1/4}.
\end{equation}
At the radius $R_{\rm E}$, $\omega_B/\omega =a$. 
\item The intermediate transition radius $R_{\rm \theta_B}$: 
As shown in \S\ref{sec:cross-section}, for $\theta'_B < 1$, the scattering cross-section for $e^\pm$ falls below $\sigma_{\rm T}$ when $\omega'_B/\omega' > a\theta'_B$ (Eq.(\ref{eq:transition})). Thus, one can define another transition radius $R_{\theta_B}$ at which $\omega'_B/\omega' = a\theta'_B$ is satisfied. The radius $R_{\theta_B}$ can be calculated as
\begin{equation}
\begin{aligned}
R_{\theta_B}&\simeq(4.2\times10^9\,{\rm cm})\; {\cal D}^{1/2}{\theta_{B,-2}'^{-1/2}} B_{\star 15}^{1/2} R_{\star,6}^{3/2} L_{\rm frb,42}^{-1/4}\\
\end{aligned}
\end{equation}

\item The outer transition radius $R_\omega$: As shown in \S\ref{sec:cross-section} the scattering cross-section becomes of the order of $\sigma_{\rm T}$ when the wave frequency ($\omega$) is larger than the cyclotron frequency of the background magnetic field ($\omega_B$). This critical radius at which the transition $\omega = \omega_B$ occurs is
\begin{equation}
R_{\omega} = \left( \frac{q B_\star R_{\star}^3 }{m_e c\omega}\right)^{1/3} = (2\times10^{10}\,{\rm cm})\; B_{\star,15}^{1/3}R_{\star,6}\nu_9^{-1/3},
\end{equation}
provided that $R_\omega$ is above the light cylinder radius $R_{\rm lc}$.
\item The light cylinder 
\begin{equation}
    R_{\rm lc} = c / \Omega = (4.8 \times 10^9 \ {\rm cm}) P
\end{equation}
defines the outer boundary of the magnetosphere.
\end{itemize}

\begin{figure}
	\includegraphics[width=\columnwidth]{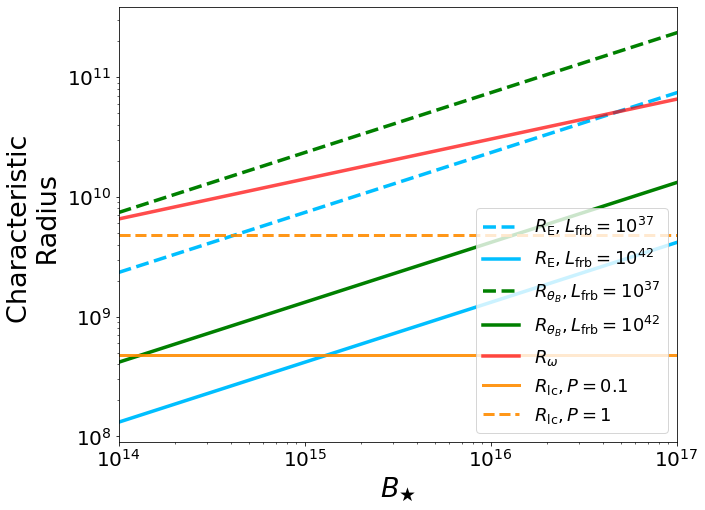}
    \caption{The four characteristic radii as a function of NS surface magnetic field $B_{\star}$, rotation period $P$, and FRB luminosity $L_{\rm frb}$. The red solid line $R_{\omega}$ is at a constant FRB frequency $\nu=1 \ \rm GHz$. $\theta_B$ and $\gamma_p$ are normalized to 0.1 and $10^3$, respectively. The dashed and solid lines denote the $R_{\rm E}$ (blue) and $R_{\theta_B}$ (green) for different FRB luminosity, i.e. $L_{\rm frb}=10^{37} \ {\rm and} \ 10^{42} \ \rm erg \ s^{-1}$, respectively. The orange dashed and solid lines denote the light cylinder radius for $P=1, 0.1 \ \rm s$, respectively. We only focus on the region where ${{\rm max}(R_{\rm E},R_{\theta_B})}\leq R\leq{\rm min}\{R_\omega,R_{\rm lc}\}$. For the parameters adopted in this figure, it corresponds to the region above the green solid line but below the orange dashed line.}
    \label{R-B}
\end{figure}

The scattering optical depth experienced by an FRB generated in a magnetar magnetosphere can be generally written as 
\begin{equation}\label{eq:tau}
\begin{aligned}
\tau&=\int_{0}^{r_{\rm lc}} n_e (1-\beta\cos\theta_B) {\sigma'} ds \\
&=\int_{0}^{r_{\rm lc}} \xi n_{\rm GJ} \sigma'\frac{(1-\beta \cos\theta_B)^2}{\cos\theta_B}dr.
\end{aligned}
\end{equation}
where 
\begin{equation}
ds=\frac{1-\beta\cos\theta_B}{\cos\theta_B}dr 
\end{equation}
denotes the increment distance of a photon travels within the plasma that is moving with a Lorentz factor $\gamma_p$ and along a direction $\theta_B$ with respect to the line of sight, $n_{\rm GJ}$ is the Goldreich Julian number density, $r_{\rm lc}$ is the distance from the FRB emission position to the light cylinder, $\xi$ is the pair multiplicity so that $\xi n_{\rm GJ}$ is the total particle number density in the magnetosphere (which is $r$-dependent), and $\sigma'$ is the comoving-frame scattering cross section that was calculated in \S\ref{sec:cross-section}. 

The region of the magnetosphere where significant scattering occurs is determined by the ordering of the transition radii we defined at the beginning of this section. To a very good approximation the integral may be written as 
\begin{equation}
\begin{aligned}
\tau &=\int_{0}^{r_{\rm lc}} \xi n_{\rm GJ} \sigma'\frac{(1-\beta \cos\theta_B)^2}{\cos\theta_B}dr\\
&\simeq\int_{R_{\rm FRB}}^{R_{\rm lc}} \xi n_{\rm GJ} \sigma'\frac{(1-\beta \cos\theta_B)^2}{\cos\theta_B}dR\\
&\simeq \int_{R_{\rm min}}^{R_{\rm max}} \xi n_{\rm GJ} \sigma'\frac{(1-\beta \cos\theta_B)^2}{\cos\theta_B}dR,
\label{eq:tau2}
\end{aligned}
\end{equation}
where the first approximation makes use a geometric relation, and the second relation only includes the radius range where the cross section is large. Here $R_{\rm min} = {\rm max} (R_{\rm FRB}, R_{\rm E}, R_{\theta_B})$ and $R_{\rm max} = {\rm min} (R_{\rm lc}, R_\omega)$. 

The dependences of $R_{\rm E}$, $R_\omega$, $R_{\theta_B}$ and $R_{\rm lc}$ on the magnetar surface magnetic field $B_\star$ and FRB luminosity $L_{\rm frb}$ are shown in Fig. \ref{R-B} for $P=1$ s and $\theta_B = 0.1$. From this plot, one can immediately see that magnetar magnetospheres are transparent to some FRB radio emission. For example, low-luminosity FRBs have $R_{\theta_{B}}$ above $R_{\rm lc}$ (for $P \geq 1$ s) so that they are transparent. Another case is that  a high luminosity FRB with $L_{\rm frb} \sim 10^{42}$ erg s$^{-1}$ can easily escape the magnetosphere if $B_\star>10^{16}$ G and $P<0.5$ s, because  $R_{\theta_B} > R_{\rm E} \geq \min (R_\omega, R_{\rm lc})$ in this case. In fact, as we show below, the radio waves of this luminous FRB can escape intact for a large range of magnetospheric parameters.

For more general cases, one needs to perform numerical integration in Equation (\ref{eq:tau}). Performing such an integration is computationally very time consuming because it involves calculating the cross-section ($\sigma$) at a large number of points along the FRB pulse trajectory, and the calculation of $\sigma$ requires calculating particle trajectories accurately for the time duration much longer than the wave period. However, we can make an approximation that drastically reduces the computation time without sacrificing the accuracy too much.  Since the integrand of Eq.(\ref{eq:tau2}) is a rapidly decreasing function of $r$, it is reasonable to calculate the integrand at $R_{\rm min}$ and approximate the expression of the optical depth as
\begin{equation}
\tau \sim \xi n_{\rm GJ} \sigma'\frac{(1-\beta \cos\theta_B)^2}{\cos\theta_B} {\rm max}(R_{\rm E},R_{\theta_B}).
\label{eq:tau3}
\end{equation}
{ When $\gamma_p \gg 1$ and $\theta_B \ll 1$, one gets
\begin{equation}
\tau \sim \xi n_{\rm GJ} \sigma'  (\theta_B^4 / 4) {\rm max}(R_{\rm E},R_{\theta_B}).
\end{equation}
In the case of $R_{\theta_B}>R_E$, one has  $n_{\rm GJ}(R_{\theta_B})\sim(10^4 \ {\rm cm^{-3}}) \ B_{\star,15}R_{\star,6}^3P^{-1}r_{9.28}^{-3}$,  $a(R_{\theta_B})\simeq8.5\times10^3 \ L_{\rm frb,42}^{1/2}\nu_9^{-1}r_{9.28}^{-1}$ (where $r$ is normalized to the value of $R_{\theta_B} \sim 10^{9.28}$ cm for our nominal parameters). We find
\begin{equation}
\tau \sim2.3\times10^{-6} \ \xi_2 n_{\rm GJ,4} \frac{\sigma'}{a(R_{\theta_B})^2\sigma_{\rm T}}  \theta_{B,-1}^4 R_{\theta_B,9.28} \ll 1.
\end{equation}
Here $\sigma'$ is normalized to $[a(R_{\theta_B})]^2 \sigma_{\rm T} \sim(4.8\times10^{-17} \ {\rm cm^2}) \ a_{3.93}^2$ for the rough estimation.

When $R_{\theta_B}<R_{\rm E}$, one has $n_{\rm GJ}(R_{\rm E})\sim(1.5\times10^5 \ {\rm cm^{-3}}) \ B_{\star,15}R_{\star,6}^3P^{-1}r_{8.8}^{-3}$ and $a(R_{\rm E})\simeq2.1\times10^4 \ L_{\rm frb,42}^{1/2}\nu_9^{-1}r_{8.8}^{-1}$, which gives
\begin{equation}
\tau\sim8.6\times10^{-5} \ \xi_2 n_{\rm GJ,4} \frac{\sigma'}{a(R_{\rm E})^2\sigma_{\rm T}}  \theta_{B,-1}^4 R_{{\rm E},8.8} < 1.
\end{equation}
Here $\sigma'$ is normalized to $[a(R_{\rm E})]^2 \sigma_{\rm T} \sim(2.9\times10^{-16} \ {\rm cm^2}) \ a_{4.32}^2$. We see that $\tau < 1$ is satisfied for the typical parameters considered in magnetispheric FRB models. 
Making use of the scaling relation Equation (\ref{eq:sigma-thetaB}) obtained from numerical results, one can also derive the following dependences
\begin{equation}
\tau \propto {\theta_B'^{3.5}}  \theta_B^4
\propto \gamma_p^{-3.5} \theta_B^{0.5}
\label{eq:tau4}
\end{equation}
for $R_{\theta_B}<R_{\rm E}$, and
\begin{equation}
\tau\propto{\theta_B'^{3.5}}  \theta_B^4\theta_B'^{-1/2}\gamma_p^{-1/2}\theta_B^{-1}\propto\gamma_p^{-3.5}
\label{eq:tau5}
\end{equation}
for $R_{\theta_B}>R_{\rm E}$. Both scalings suggest that FRBs tend to become transparent in the large $\gamma_p$ and small $\theta_B$ regimes. As argued in \S\ref{sec:streaming} and \S\ref{sec:thetaB}, these two regimes are relevant for FRB magnetospheric emission models.}

\begin{figure}
	\includegraphics[width=\columnwidth]{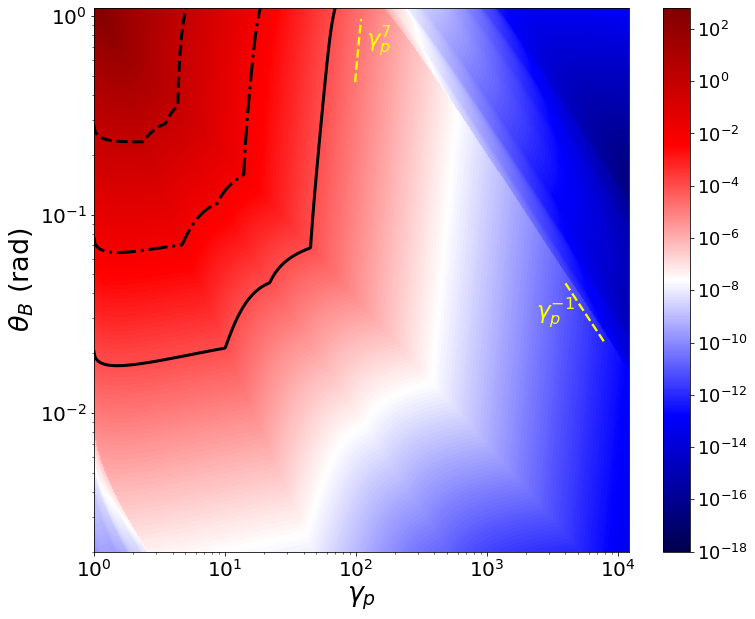}
    \caption{The optical depth ($\tau$) as a function of plasma Lorentz factor $\gamma_p$ and $\theta_B$ (units in radian). Following parameters are adopted: $L_{\rm frb}=10^{42} \ {\rm erg \ s^{-1}}$, $\nu=1 \ \rm GHz$, $R={\rm max}(R_{\rm E},R_{\theta_B})$, $\xi=1$, $a=10^4$ and $\omega_B'/\omega'=35$. The dashed, dashed-dotted and solid black lines mark the optical depth $\tau=1, 10^{-2}$ and $10^{-4}$, which is equivalently $\tau=1$ for different multiplicity parameters $\xi=1, 100$ and $\xi=10^4$, respectively. {The yellow dashed lines are reference lines with slopes 7 and -1, respectively}.}
    \label{contour}
\end{figure}

In Fig. \ref{contour}, we present the numerical results of the scattering optical depth ($\tau$) as a function of $\gamma_p$ and $\theta_B$. Because of the large computation time to calculate cross sections for different $\theta'_B$ and $\omega/\omega_B$ parameters, we did not directly perform the integration in Eq.(\ref{eq:tau2}). Rather, we use the approximate formula Eq.(\ref{eq:tau3}), which is sufficiently accurate for determining whether FRBs can escape the magnetosphere of a magnetar. 
In our calculation, $\gamma_p$ and $\theta_B$ are adopted as free parameters. Other parameters include $L_{\rm frb}=10^{42} \ {\rm erg \ s^{-1}}$ at $\nu=1 \ \rm GHz$, $\xi=1$, $R={\rm max}(R_{\rm E},R_{\theta_B})$, and $a=10^4$. We adopt these nominal values where the scattering cross section has been calculated numerically as a function of $\theta'_B$ (Fig. \ref{omega_B=35}).  The three black curves mark the contours for $\tau = 1$ (dashed), $\tau = 0.01$ (dash-dotted) and $\tau = 10^{-4}$ (solid), respectively, for $\xi=1$. Since $\tau \propto \xi$, these three curves are also the effective $\tau = 1$ curves for three multiplicity values $\xi = 1, 100, 10^4$, respectively. The adoption of these three $\xi$ values is based on the following considerations: 1. For coherent curvature radiation by bunches \citep{Kumar2017,Yang&zhang2018,Qu&Zhang21}, $\xi$ should be of order $10^2$; 2. For coherent inverse Compton scattering by bunches \citep{Zhang2022}, the required $\xi$ is much smaller and could be or order of unity; 3. \cite{Beloborodov2021} argued that pair production cascade may happen due to the large amplitude wave effect, which may increase $\xi$ substantially. However, as shown in \S\ref{contour} below, when the plasma is moving relativistically, the multiplicity $\xi$ is at most $\sim 10^4$. In any case, an FRB can escape the magnetosphere in the parameter space below and to the right of these curves. { One can see that an FRB with $L_{\rm frb} = 10^{42} \ {\rm erg \ s^{-1}}$ can escape the magnetar magnetosphere as long as $\gamma_p > 10^2$.} This is typically the case as we argued in \S\ref{sec:streaming}. At small $\theta_B$ values, the required minimum $\gamma_p$ is correspondingly smaller. As shown in Fig. \ref{theta_B}, the maximum $\theta_B$ is larger for more rapid rotators, so that a slower rotator could be more transparent to FRBs. The $\vec k$-$\vec B$ alignment effect (\S\ref{sec:alignment}) could further facilitate the transparency of FRB waves.

Some interesting features are visible in Fig. \ref{contour}, which can be understood from the analytical estimates. There is a clear edge on the right side of the diagram with a slopt of $\theta_B \propto \gamma_p^{-1}$. This arises from the fact that the scattering cross section drops rapidly for $\omega'_B/\omega' \gta a \theta'_B$, and is responsible for the scaling $\theta_B \propto \gamma_p^{-1}$ we see in the figure. { To the left of that boundary, the contour shows a rising slope 7.} This corresponds to the constant $\tau$ line as defined by Equations (\ref{eq:tau4}) and (\ref{eq:tau5}). Further to the left, the contour shape becomes more complex, which is a result of the convolution between the Doppler factor and the competition between $R_{\theta_B}$ and $R_{\rm E}$ in that region of the parameter space.

\section{Maximum pair multiplicity due to electron radiation in FRB waves} 

Electrons accelerated by the large-amplitude FRB waves will radiate a generic curvature radiation in their curved trajectories and radiate $\gamma$-ray photons. These photons may produce additional $e^\pm$, which would further increase the optical depth for scattering off the FRB waves thereby preventing the escape of FRB pulse from the NS magnetosphere \citep{Beloborodov2021}. Assume that the curvature radius is of the order of $\rho = (10^9 \ {\rm cm}) \ \rho_9$. The characteristic electron Lorentz factor required to produce photon energy of the order of $m c^2$ is
\begin{equation}
\gamma_c=\left(\frac{4\pi\rho\nu_c}{3c}\right)^{1/3}\simeq2.6\times10^6 \ \rho_9^{1/3},
\end{equation}
where $\nu_c=m_ec^2/h$.
To calculate the multiplicity $\xi$, one needs to investigate the photon number density and the mean free path of a photon before producing a pair. Here instead of going through these processes, we present a very general argument to derive the maximum of pair multiplicity $\xi$ in a plasma moving with a bulk Lorentz factor $\gamma_p$.

An extreme assumption is that the entire FRB luminosity is converted to the energy of pair-radiating plasma. This requires 
\begin{equation}
L_{\rm frb} = \dot N \gamma_p \gamma_c m_e c^2,
\end{equation}
where $\dot N$ is the isotropic-equivalent particle injection rate in the emission region, which can be related to the particle density through
\begin{equation}
n = \frac{\dot N}{4\pi r^2 c} \simeq (1.2\times10^9 \ {\rm cm^{-3}}) \ L_{\rm frb,42}\gamma_{c,6}^{-1}\gamma_{p,3}^{-1}r_9^{-2}.
\end{equation}
Compared with the Goldreich-Julian density $n_{\rm GJ}=B_\star/(ecP)(r/R_\star)^{-3}$, one can see that the maximum achievable multiplicity is 
\begin{equation}
\xi = \frac{n}{n_{\rm GJ}}\simeq1.8\times10^4 \ L_{\rm frb,42}\gamma_{c,6}^{-1}\gamma_{p,3}^{-1}r_9B_{\star,15}^{-1}PR_{\star,6}^{-3}.
\end{equation}
One can see that for our nominal parameter $\gamma_p \sim 10^2$, the maximum $\xi$ is of the order of 
$\sim 10^4$. This value is adopted as the extreme case in our optical depth calculation in Figure \ref{contour}.  One can see that even for this case, bright FRBs are transparent in magnetar magnetospheres as long as $\gamma_p$ is of the order of $10^2$ and higher.

Notice that the calculation presented in this section only applies to FRB-wave-induced pair cascade as advocated by \cite{Beloborodov2021}. In a magnetar magnetosphere, pair cascades can be triggered by other processes. Most pairs are generated near the surface, which would not affect the propagation of the FRBs emitted from large radii. However, FRBs may be associated with bright hard X-ray bursts or even giant soft-gamma-ray flares. One needs to further consider the opacity introduced by the pairs produced in these high-energy events and discuss the breakout of FRBs from these fireballs. As shown by \cite{Ioka2020}, FRBs can be choked if the high-energy bursts are too bright, e.g. in the case of giant flares.

\section{Conclusions and Discussion}\label{sec:conclusions}

In this paper, we have investigated the propagation of a nonlinear EM wave, associated with a bright FRB, through the magnetosphere of a magnetar. In particular, we have calculated the optical depth ($\tau$) of the magnetosphere to FRB waves due to their scattering by the electron-positron plasma and showed that $\tau$ is small for realistic physical conditions. Our main conclusions can be summarized as follows:

\begin{itemize}
\item For an $e^\pm$ initially at rest, we numerically solved the relativistic equation of motion of a particle under the influence of a linearly polarized FRB pulse and a background static magnetic field (\S\ref{sec:cross-section}). We confirmed the previous results that the scattering cross section is significantly enhanced when large-amplitude waves propagate through the outer magnetosphere, $r \gta 10^9$cm, where the static magnetic field strength is smaller than the wave-field and when the electron cyclotron frequency ($\omega_B$) is larger than the wave frequency ($\omega$). The average cross section, $\sigma$, sensitively depends on $\omega_B/\omega$, $\theta_B$, and $a$; where $\theta_B$ is angle between the background magnetic field and the wave propagation direction, and $a$ is the dimensionless EM wave amplitude or nonlinearity parameter. In particular, $\sigma'$ becomes much smaller than the Thomson cross-section ($\sigma_{\rm T}$) when $a \theta'_B  \lta \omega'_B/\omega'$. This makes the scattering optical depth of the magnetar outer magnetosphere to radio waves with $a\gg 1$ much less than unity under certain conditions.

\item In \S\ref{sec:streaming}, we presented a couple of arguments that the plasma in the outer magnetosphere, especially in the open field line regions, { is likely streaming outward with a bulk Lorentz factor $\gamma_p \geq 10^2$.} The arguments include the standard pulsar electrodynamics in the open field line regions, Alfv\'en wave charge depletion acceleration, and ponderomotive force acceleration of the plasma at the front end of the FRB waves. In fact, the ponderomotive force associated with a weak precursor to the observed FRB event might be sufficient to accelerate the plasma and clear the outer magnetosphere enabling the main FRB pulse to propagate unimpeded. In any case, the optical depth of plasma to nonlinear EM waves is highly reduced when it is moving away with $\gamma_p\gg 1$.

\item In \S\ref{sec:thetaB}, we showed that the angle between the wave propagation direction and magnetic field direction, $\theta_B$, is generally small in the open field line region of the outer magnetosphere when the waves are generated at $r\lta 10^8$ cm. Furthermore, for plasma moving along the background magnetic field with $\gamma_p\gg 1$, before it encounters the FRB pulse, the angle $\theta'_B$ in the plasma comoving frame is further reduced due to the relativistic aberration of light, i.e. the wave propagation direction is very nearly aligned with the static magnetic field in the plasma comoving frame. We showed in \S\ref{sec:cross-section} that the scattering cross section is greatly reduced when $\theta'_B \ll 1$. 

\item We presented in \S\ref{sec:tau} the scattering optical depth for a bright FRB with luminosity $10^{42} \ {\rm erg \ s^{-1}}$ in a magnetar magnetosphere by combining the results described above. We found that the magnetar magnetosphere is transparent to large amplitude FRB radio waves for a large region of the $\gamma_p$--$\theta_B$ parameter space.

\end{itemize}

This result, that the radio pulses of bright FRBs can pass through the magnetar magnetosphere unscathed, removes the recently suggested theoretical objection against the magnetospheric origin of GRBs, which has been otherwise supported by observational data.

\section*{Acknowledgements}
This work is partially supported by the Top Tier Doctoral Graduate Research Assistantship (TTDGRA) and Nevada Center for Astrophysics at the University of Nevada, Las Vegas. PK thanks the National Science Foundation for their support for this work through the grant AST-2009619.

%%%%%%%%%%%%%%%%%%%%%%%%%%%%%%%%%%%%%%%%%%%%%%%%%%
\section*{Data Availability}
The code developed to perform the calculation in this paper is available upon request.



%\bibliographystyle{mnras}

\begin{thebibliography}{}
\makeatletter
\relax
\def\mn@urlcharsother{\let\do\@makeother \do\$\do\&\do\#\do\^\do\_\do\%\do\~}
\def\mn@doi{\begingroup\mn@urlcharsother \@ifnextchar [ {\mn@doi@}
  {\mn@doi@[]}}
\def\mn@doi@[#1]#2{\def\@tempa{#1}\ifx\@tempa\@empty \href
  {http://dx.doi.org/#2} {doi:#2}\else \href {http://dx.doi.org/#2} {#1}\fi
  \endgroup}
\def\mn@eprint#1#2{\mn@eprint@#1:#2::\@nil}
\def\mn@eprint@arXiv#1{\href {http://arxiv.org/abs/#1} {{\tt arXiv:#1}}}
\def\mn@eprint@dblp#1{\href {http://dblp.uni-trier.de/rec/bibtex/#1.xml}
  {dblp:#1}}
\def\mn@eprint@#1:#2:#3:#4\@nil{\def\@tempa {#1}\def\@tempb {#2}\def\@tempc
  {#3}\ifx \@tempc \@empty \let \@tempc \@tempb \let \@tempb \@tempa \fi \ifx
  \@tempb \@empty \def\@tempb {arXiv}\fi \@ifundefined
  {mn@eprint@\@tempb}{\@tempb:\@tempc}{\expandafter \expandafter \csname
  mn@eprint@\@tempb\endcsname \expandafter{\@tempc}}}

\bibitem[\protect\citeauthoryear{{Arons} \& {Scharlemann}}{{Arons} \&
  {Scharlemann}}{1979}]{Arons79}
{Arons} J.,  {Scharlemann} E.~T.,  1979, \mn@doi [\apj] {10.1086/157250}, \href
  {https://ui.adsabs.harvard.edu/abs/1979ApJ...231..854A} {231, 854}

\bibitem[\protect\citeauthoryear{{Baring} \& {Harding}}{{Baring} \&
  {Harding}}{1998}]{baring98}
{Baring} M.~G.,  {Harding} A.~K.,  1998, \mn@doi [\apjl] {10.1086/311679},
  \href {https://ui.adsabs.harvard.edu/abs/1998ApJ...507L..55B} {507, L55}

\bibitem[\protect\citeauthoryear{{Bauer}, {Mulser}  \& {Steeb}}{{Bauer}
  et~al.}{1995}]{Bauer1995}
{Bauer} D.,  {Mulser} P.,   {Steeb} W.~H.,  1995, \mn@doi [\prl]
  {10.1103/PhysRevLett.75.4622}, \href
  {https://ui.adsabs.harvard.edu/abs/1995PhRvL..75.4622B} {75, 4622}

\bibitem[\protect\citeauthoryear{{Beloborodov}}{{Beloborodov}}{2020}]{Beloborodov2020}
{Beloborodov} A.~M.,  2020, \mn@doi [\apj] {10.3847/1538-4357/ab83eb}, \href
  {https://ui.adsabs.harvard.edu/abs/2020ApJ...896..142B} {896, 142}

\bibitem[\protect\citeauthoryear{{Beloborodov}}{{Beloborodov}}{2021}]{Beloborodov2021}
{Beloborodov} A.~M.,  2021, \mn@doi [\apjl] {10.3847/2041-8213/ac2fa0}, \href
  {https://ui.adsabs.harvard.edu/abs/2021ApJ...922L...7B} {922, L7}

\bibitem[\protect\citeauthoryear{{Bochenek}, {Ravi}, {Belov}, {Hallinan},
  {Kocz}, {Kulkarni}  \& {McKenna}}{{Bochenek} et~al.}{2020}]{Bochenek2020}
{Bochenek} C.~D.,  {Ravi} V.,  {Belov} K.~V.,  {Hallinan} G.,  {Kocz} J.,
  {Kulkarni} S.~R.,   {McKenna} D.~L.,  2020, \mn@doi [\nat]
  {10.1038/s41586-020-2872-x}, \href
  {https://ui.adsabs.harvard.edu/abs/2020Natur.587...59B} {587, 59}

\bibitem[\protect\citeauthoryear{{CHIME/FRB Collaboration} et~al.,}{{CHIME/FRB
  Collaboration} et~al.}{2019}]{CHIME19}
{CHIME/FRB Collaboration} et~al., 2019, \mn@doi [\apjl]
  {10.3847/2041-8213/ab4a80}, \href
  {https://ui.adsabs.harvard.edu/abs/2019ApJ...885L..24C} {885, L24}

\bibitem[\protect\citeauthoryear{{CHIME/FRB Collaboration} et~al.,}{{CHIME/FRB
  Collaboration} et~al.}{2020}]{CHIME/FRB2020}
{CHIME/FRB Collaboration} et~al., 2020, \mn@doi [\nat]
  {10.1038/s41586-020-2863-y}, \href
  {https://ui.adsabs.harvard.edu/abs/2020Natur.587...54C} {587, 54}

\bibitem[\protect\citeauthoryear{{Chen}, {Yuan}, {Beloborodov}  \& {Li}}{{Chen}
  et~al.}{2022}]{chen2022}
{Chen} A.~Y.,  {Yuan} Y.,  {Beloborodov} A.~M.,   {Li} X.,  2022, \mn@doi
  [\apj] {10.3847/1538-4357/ac59b1}, \href
  {https://ui.adsabs.harvard.edu/abs/2022ApJ...929...31C} {929, 31}

\bibitem[\protect\citeauthoryear{{Daugherty} \& {Harding}}{{Daugherty} \&
  {Harding}}{1996}]{daugherty96}
{Daugherty} J.~K.,  {Harding} A.~K.,  1996, \mn@doi [\apj] {10.1086/176811},
  \href {https://ui.adsabs.harvard.edu/abs/1996ApJ...458..278D} {458, 278}

\bibitem[\protect\citeauthoryear{{Goldreich} \& {Julian}}{{Goldreich} \&
  {Julian}}{1969}]{GJ1969}
{Goldreich} P.,  {Julian} W.~H.,  1969, \mn@doi [\apj] {10.1086/150119}, \href
  {https://ui.adsabs.harvard.edu/abs/1969ApJ...157..869G} {157, 869}

\bibitem[\protect\citeauthoryear{{Hilmarsson}, {Spitler}, {Main}  \&
  {Li}}{{Hilmarsson} et~al.}{2021}]{Hilmarsson21}
{Hilmarsson} G.~H.,  {Spitler} L.~G.,  {Main} R.~A.,   {Li} D.~Z.,  2021,
  \mn@doi [\mnras] {10.1093/mnras/stab2936}, \href
  {https://ui.adsabs.harvard.edu/abs/2021MNRAS.508.5354H} {508, 5354}

\bibitem[\protect\citeauthoryear{{Ioka}}{{Ioka}}{2020}]{Ioka2020}
{Ioka} K.,  2020, \mn@doi [\apjl] {10.3847/2041-8213/abc6a3}, \href
  {https://ui.adsabs.harvard.edu/abs/2020ApJ...904L..15I} {904, L15}

\bibitem[\protect\citeauthoryear{{Jackson}}{{Jackson}}{1998}]{Jackson1998}
{Jackson} J.~D.,  1998, {Classical Electrodynamics, 3rd Edition}

\bibitem[\protect\citeauthoryear{{Kumar} \& {Bo{\v{s}}njak}}{{Kumar} \&
  {Bo{\v{s}}njak}}{2020}]{Kumar&Bosnjak2020}
{Kumar} P.,  {Bo{\v{s}}njak} {\v{Z}}.,  2020, \mn@doi [\mnras]
  {10.1093/mnras/staa774}, \href
  {https://ui.adsabs.harvard.edu/abs/2020MNRAS.494.2385K} {494, 2385}

\bibitem[\protect\citeauthoryear{{Kumar} \& {Lu}}{{Kumar} \&
  {Lu}}{2020}]{Kumar&Lu2020}
{Kumar} P.,  {Lu} W.,  2020, \mn@doi [\mnras] {10.1093/mnras/staa801}, \href
  {https://ui.adsabs.harvard.edu/abs/2020MNRAS.494.1217K} {494, 1217}

\bibitem[\protect\citeauthoryear{{Kumar}, {Lu}  \& {Bhattacharya}}{{Kumar}
  et~al.}{2017}]{Kumar2017}
{Kumar} P.,  {Lu} W.,   {Bhattacharya} M.,  2017, \mn@doi [\mnras]
  {10.1093/mnras/stx665}, \href
  {https://ui.adsabs.harvard.edu/abs/2017MNRAS.468.2726K} {468, 2726}

\bibitem[\protect\citeauthoryear{{Li} et~al.,}{{Li} et~al.}{2021a}]{Li21}
{Li} C.~K.,  et~al., 2021a, \mn@doi [Nature Astronomy]
  {10.1038/s41550-021-01302-6}, \href
  {https://ui.adsabs.harvard.edu/abs/2021NatAs...5..378L} {5, 378}

\bibitem[\protect\citeauthoryear{{Li} et~al.,}{{Li} et~al.}{2021b}]{LiD2021}
{Li} D.,  et~al., 2021b, \mn@doi [\nat] {10.1038/s41586-021-03878-5}, \href
  {https://ui.adsabs.harvard.edu/abs/2021Natur.598..267L} {598, 267}

\bibitem[\protect\citeauthoryear{{Lin} et~al.,}{{Lin} et~al.}{2020}]{Lin20}
{Lin} L.,  et~al., 2020, \mn@doi [\nat] {10.1038/s41586-020-2839-y}, \href
  {https://ui.adsabs.harvard.edu/abs/2020Natur.587...63L} {587, 63}

\bibitem[\protect\citeauthoryear{{Lorimer}, {Bailes}, {McLaughlin}, {Narkevic}
  \& {Crawford}}{{Lorimer} et~al.}{2007}]{Lorimer07}
{Lorimer} D.~R.,  {Bailes} M.,  {McLaughlin} M.~A.,  {Narkevic} D.~J.,
  {Crawford} F.,  2007, \mn@doi [Science] {10.1126/science.1147532}, \href
  {https://ui.adsabs.harvard.edu/abs/2007Sci...318..777L} {318, 777}

\bibitem[\protect\citeauthoryear{{Lu}, {Kumar}  \& {Zhang}}{{Lu}
  et~al.}{2020}]{Lu20}
{Lu} W.,  {Kumar} P.,   {Zhang} B.,  2020, \mn@doi [\mnras]
  {10.1093/mnras/staa2450}, \href
  {https://ui.adsabs.harvard.edu/abs/2020MNRAS.498.1397L} {498, 1397}

\bibitem[\protect\citeauthoryear{{Luan} \& {Goldreich}}{{Luan} \&
  {Goldreich}}{2014}]{Luan2014}
{Luan} J.,  {Goldreich} P.,  2014, \mn@doi [\apjl]
  {10.1088/2041-8205/785/2/L26}, \href
  {https://ui.adsabs.harvard.edu/abs/2014ApJ...785L..26L} {785, L26}

\bibitem[\protect\citeauthoryear{{Luo} et~al.,}{{Luo} et~al.}{2020}]{Luo2020}
{Luo} R.,  et~al., 2020, \mn@doi [\nat] {10.1038/s41586-020-2827-2}, \href
  {https://ui.adsabs.harvard.edu/abs/2020Natur.586..693L} {586, 693}

\bibitem[\protect\citeauthoryear{{Lyubarsky}}{{Lyubarsky}}{2014}]{Lyubarsky2014}
{Lyubarsky} Y.,  2014, \mn@doi [\mnras] {10.1093/mnrasl/slu046}, \href
  {https://ui.adsabs.harvard.edu/abs/2014MNRAS.442L...9L} {442, L9}

\bibitem[\protect\citeauthoryear{{Lyubarsky}}{{Lyubarsky}}{2020}]{Lyubarsky2020}
{Lyubarsky} Y.,  2020, \mn@doi [\apj] {10.3847/1538-4357/ab97b5}, \href
  {https://ui.adsabs.harvard.edu/abs/2020ApJ...897....1L} {897, 1}

\bibitem[\protect\citeauthoryear{{Margalit}, {Beniamini}, {Sridhar}  \&
  {Metzger}}{{Margalit} et~al.}{2020}]{Margalit20}
{Margalit} B.,  {Beniamini} P.,  {Sridhar} N.,   {Metzger} B.~D.,  2020,
  \mn@doi [\apjl] {10.3847/2041-8213/abac57}, \href
  {https://ui.adsabs.harvard.edu/abs/2020ApJ...899L..27M} {899, L27}

\bibitem[\protect\citeauthoryear{{Mereghetti} et~al.,}{{Mereghetti}
  et~al.}{2020}]{Mereghetti20}
{Mereghetti} S.,  et~al., 2020, \mn@doi [\apjl] {10.3847/2041-8213/aba2cf},
  \href {https://ui.adsabs.harvard.edu/abs/2020ApJ...898L..29M} {898, L29}

\bibitem[\protect\citeauthoryear{{Metzger}, {Margalit}  \& {Sironi}}{{Metzger}
  et~al.}{2019}]{Metzger2019}
{Metzger} B.~D.,  {Margalit} B.,   {Sironi} L.,  2019, \mn@doi [\mnras]
  {10.1093/mnras/stz700}, \href
  {https://ui.adsabs.harvard.edu/abs/2019MNRAS.485.4091M} {485, 4091}

\bibitem[\protect\citeauthoryear{{Muslimov} \& {Tsygan}}{{Muslimov} \&
  {Tsygan}}{1992}]{Muslimov92}
{Muslimov} A.~G.,  {Tsygan} A.~I.,  1992, \mn@doi [\mnras]
  {10.1093/mnras/255.1.61}, \href
  {https://ui.adsabs.harvard.edu/abs/1992MNRAS.255...61M} {255, 61}

\bibitem[\protect\citeauthoryear{{Qiao} \& {Lin}}{{Qiao} \&
  {Lin}}{1998}]{Qiao&Lin1998}
{Qiao} G.~J.,  {Lin} W.~P.,  1998, \aap, \href
  {https://ui.adsabs.harvard.edu/abs/1998A&A...333..172Q} {333, 172}

\bibitem[\protect\citeauthoryear{{Qu} \& {Zhang}}{{Qu} \&
  {Zhang}}{2021}]{Qu&Zhang21}
{Qu} Y.,  {Zhang} B.,  2021, arXiv e-prints, \href
  {https://ui.adsabs.harvard.edu/abs/2021arXiv211112269Q} {p. arXiv:2111.12269}

\bibitem[\protect\citeauthoryear{{Ruderman} \& {Sutherland}}{{Ruderman} \&
  {Sutherland}}{1975}]{ruderman75}
{Ruderman} M.~A.,  {Sutherland} P.~G.,  1975, \mn@doi [\apj] {10.1086/153393},
  \href {https://ui.adsabs.harvard.edu/abs/1975ApJ...196...51R} {196, 51}

\bibitem[\protect\citeauthoryear{{Sironi}, {Plotnikov}, {N{\"a}ttil{\"a}}  \&
  {Beloborodov}}{{Sironi} et~al.}{2021}]{Sironi2021}
{Sironi} L.,  {Plotnikov} I.,  {N{\"a}ttil{\"a}} J.,   {Beloborodov} A.~M.,
  2021, \mn@doi [\prl] {10.1103/PhysRevLett.127.035101}, \href
  {https://ui.adsabs.harvard.edu/abs/2021PhRvL.127c5101S} {127, 035101}

\bibitem[\protect\citeauthoryear{{Thompson}}{{Thompson}}{2008}]{thompson08}
{Thompson} C.,  2008, \mn@doi [\apj] {10.1086/592061}, \href
  {https://ui.adsabs.harvard.edu/abs/2008ApJ...688..499T} {688, 499}

\bibitem[\protect\citeauthoryear{{Thompson} \& {Duncan}}{{Thompson} \&
  {Duncan}}{1995}]{Thompson95}
{Thompson} C.,  {Duncan} R.~C.,  1995, \mn@doi [\mnras]
  {10.1093/mnras/275.2.255}, \href
  {https://ui.adsabs.harvard.edu/abs/1995MNRAS.275..255T} {275, 255}

\bibitem[\protect\citeauthoryear{{Thompson} \& {Duncan}}{{Thompson} \&
  {Duncan}}{1996}]{Thompson96}
{Thompson} C.,  {Duncan} R.~C.,  1996, \mn@doi [\apj] {10.1086/178147}, \href
  {https://ui.adsabs.harvard.edu/abs/1996ApJ...473..322T} {473, 322}

\bibitem[\protect\citeauthoryear{{Thornton} et~al.,}{{Thornton}
  et~al.}{2013}]{Thornton13}
{Thornton} D.,  et~al., 2013, \mn@doi [Science] {10.1126/science.1236789},
  \href {https://ui.adsabs.harvard.edu/abs/2013Sci...341...53T} {341, 53}

\bibitem[\protect\citeauthoryear{{Usov}}{{Usov}}{2002}]{usov02}
{Usov} V.~V.,  2002, \mn@doi [\apjl] {10.1086/341505}, \href
  {https://ui.adsabs.harvard.edu/abs/2002ApJ...572L..87U} {572, L87}

\bibitem[\protect\citeauthoryear{{Wadiasingh}, {Beniamini}, {Timokhin},
  {Baring}, {van der Horst}, {Harding}  \& {Kazanas}}{{Wadiasingh}
  et~al.}{2020}]{Wadiasingh20}
{Wadiasingh} Z.,  {Beniamini} P.,  {Timokhin} A.,  {Baring} M.~G.,  {van der
  Horst} A.~J.,  {Harding} A.~K.,   {Kazanas} D.,  2020, \mn@doi [\apj]
  {10.3847/1538-4357/ab6d69}, \href
  {https://ui.adsabs.harvard.edu/abs/2020ApJ...891...82W} {891, 82}

\bibitem[\protect\citeauthoryear{{Wang}, {Yang}, {Niu}, {Xu}  \&
  {Zhang}}{{Wang} et~al.}{2022}]{Wang2022}
{Wang} W.-Y.,  {Yang} Y.-P.,  {Niu} C.-H.,  {Xu} R.,   {Zhang} B.,  2022,
  \mn@doi [\apj] {10.3847/1538-4357/ac4097}, \href
  {https://ui.adsabs.harvard.edu/abs/2022ApJ...927..105W} {927, 105}

\bibitem[\protect\citeauthoryear{{Xu} et~al.,}{{Xu} et~al.}{2021}]{Xu21}
{Xu} H.,  et~al., 2021, arXiv e-prints, \href
  {https://ui.adsabs.harvard.edu/abs/2021arXiv211111764X} {p. arXiv:2111.11764}

\bibitem[\protect\citeauthoryear{{Yang} \& {Zhang}}{{Yang} \&
  {Zhang}}{2018}]{Yang&zhang2018}
{Yang} Y.-P.,  {Zhang} B.,  2018, \mn@doi [\apj] {10.3847/1538-4357/aae685},
  \href {https://ui.adsabs.harvard.edu/abs/2018ApJ...868...31Y} {868, 31}

\bibitem[\protect\citeauthoryear{{Yang} \& {Zhang}}{{Yang} \&
  {Zhang}}{2020}]{Yang&Zhang2020}
{Yang} Y.-P.,  {Zhang} B.,  2020, \mn@doi [\apjl] {10.3847/2041-8213/ab7ccf},
  \href {https://ui.adsabs.harvard.edu/abs/2020ApJ...892L..10Y} {892, L10}

\bibitem[\protect\citeauthoryear{{Yang} \& {Zhang}}{{Yang} \&
  {Zhang}}{2021}]{Yang&Zhang21}
{Yang} Y.-P.,  {Zhang} B.,  2021, \mn@doi [\apj] {10.3847/1538-4357/ac14b5},
  \href {https://ui.adsabs.harvard.edu/abs/2021ApJ...919...89Y} {919, 89}

\bibitem[\protect\citeauthoryear{{Zhang}}{{Zhang}}{2001}]{zhang2001}
{Zhang} B.,  2001, \mn@doi [\apjl] {10.1086/338051}, \href
  {https://ui.adsabs.harvard.edu/abs/2001ApJ...562L..59Z} {562, L59}

\bibitem[\protect\citeauthoryear{{Zhang}}{{Zhang}}{2003}]{zhang2002}
{Zhang} B.,  2003, Acta Astronomica Sinica, \href
  {https://ui.adsabs.harvard.edu/abs/2003AcASn..44S.215Z} {44, 215}

\bibitem[\protect\citeauthoryear{{Zhang}}{{Zhang}}{2020}]{Zhang2020}
{Zhang} B.,  2020, \mn@doi [\nat] {10.1038/s41586-020-2828-1}, \href
  {https://ui.adsabs.harvard.edu/abs/2020Natur.587...45Z} {587, 45}

\bibitem[\protect\citeauthoryear{{Zhang}}{{Zhang}}{2022}]{Zhang2022}
{Zhang} B.,  2022, \mn@doi [\apj] {10.3847/1538-4357/ac3979}, \href
  {https://ui.adsabs.harvard.edu/abs/2022ApJ...925...53Z} {925, 53}

\bibitem[\protect\citeauthoryear{{Zhang}, {Harding}  \& {Muslimov}}{{Zhang}
  et~al.}{2000}]{Zhang2000}
{Zhang} B.,  {Harding} A.~K.,   {Muslimov} A.~G.,  2000, \mn@doi [\apjl]
  {10.1086/312542}, \href
  {https://ui.adsabs.harvard.edu/abs/2000ApJ...531L.135Z} {531, L135}

\makeatother
\end{thebibliography}

\appendix

\section{Transformation of scattering cross section from co-moving frame to lab frame}\label{A}
In this appendix, we derive the transformation of scattering cross section from co-moving frame to lab frame.
We consider the photon number flux ${\partial n_\gamma}/{\partial t\partial A}$, where $A$ is the area of incident photons. The scattering cross section in the co-moving frame is defined as
\begin{equation}
\sigma'=\left<\frac{h\nu_{\rm out}'}{h\nu_{\rm in}'}\right>\frac{dn_{\rm out}/dt'}{\partial n_\gamma'/(\partial t' \partial A')},
\end{equation}
where $dn_{\rm out}/dt'$ denotes the total scattered photon number density per unit time and $h$ is the Planck constant. Notice $n_{\rm out}$ is the total number at the time duration $dt$ thus it's a Lorentz invariant. For convenience, we assume $f=h\nu_{\rm in}\partial n_\gamma/(\partial t\partial A)$ as the incident wave energy flux. We do the Doppler transformation for every physical parameter below
\begin{equation}
\begin{aligned} 
\sigma&=\left<\frac{h\nu_{\rm out}'}{h\nu_{\rm in}'}\right> \frac{dn_{\rm out}/dt}{\partial n_\gamma/(\partial t \partial A)}\\
&=\left<\frac{h\nu_{\rm out}'}{h\nu_{\rm in}'}\right>\frac{hdn_{\rm out}/dt'}{f'/\nu_{\rm in}'}\frac{dt'/dt}{f/f'}\frac{\nu_{\rm in}}{\nu_{\rm in}'}\\
&=\left<\frac{h\nu_{\rm out}'}{h\nu_{\rm in}'}\right>\frac{dn_{\rm out}/dt'}{\partial n_\gamma'/(\partial t' \partial A')}(1-\beta\cos\theta_B) \\
&=\sigma'(1-\beta\cos\theta_B),
\end{aligned}
\end{equation}
where we have applied $dt'/dt=1/\gamma$, $f/f'={\cal D}^2$ and $\nu_{\rm in}/\nu_{\rm in}'=\cal D$.

\bsp	% typesetting comment
\label{lastpage}
\end{document}